\documentclass[aps,prd,amsmath,preprint]{revtex4}

\usepackage[utf8]{inputenc}
\usepackage{amsmath,amssymb}
\usepackage{bm}
\usepackage{amscd}	% for extensible arrows (e.g., limits)
\usepackage{graphicx}
\usepackage{dcolumn}
\usepackage{url}
\usepackage{color}
\usepackage[colorlinks=true,linkcolor=blue]{hyperref}

\preprint{JLAB-THY-13-1810}

\begin{document}

\title{Impact of hadronic and nuclear corrections on	\\
	global analysis of spin-dependent parton distributions}

\author{P.~Jimenez-Delgado$^1$,
        A.~Accardi$^{1,2}$,
	W.~Melnitchouk$^1$}
\affiliation{
$^1$\mbox{Jefferson Lab, Newport News, Virginia 23606, USA}	\\
$^2$\mbox{Hampton University, Hampton, Virginia 23668, USA}	\\
{\bf Jefferson Lab Angular Momentum (JAM) Collaboration}\\
}

\date{\today\\}

\begin{abstract}
We present the first results of a new global next-to-leading order
analysis of spin-dependent parton distribution functions from the most
recent world data on inclusive polarized deep-inelastic scattering,
focusing in particular on the large-$x$ and low-$Q^2$ regions.
By directly fitting polarization asymmetries we eliminate biases
introduced by using polarized structure function data extracted
under nonuniform assumptions for the unpolarized structure functions.
For analysis of the large-$x$ data we implement nuclear smearing
corrections for deuterium and $^3$He nuclei, and systematically
include target mass and higher twist corrections to the $g_1$
and $g_2$ structure functions at low $Q^2$.
We also explore the effects of $Q^2$ and $W^2$ cuts in the data sets,
and the potential impact of future data on the behavior of the
spin-dependent parton distributions at intermediate and large $x$.
\end{abstract}

\maketitle

%%%%%%%%%%%%%%%%%%%%%%%%%%%%%%%%%%%%%%%%%%%%%%%%%%%%%%%%%%%%%%%%%%%%%%%%%
\section{Introduction}
\label{sec:intro}

The decomposition of the nucleon's spin into its constituent quark
and gluon components remains one of the most important challenges in
nuclear and particle physics.  Significant progress on this problem
has been made over the last 25 years, since the early polarized
deep-inelastic scattering (DIS) experiments \cite{EMC89} indicated
that quarks carry only a small fraction of the proton's spin.
In the intervening years a number of dedicated experimental programs
have been undertaken at various accelerator facilities worldwide,
refining our knowledge of the spin distributions through measurements
of polarization asymmetries in inclusive and semi-inclusive
lepton--nucleon scattering and polarized proton--proton collisions
(for recent reviews, see Refs.~\cite{Kuhn09, Aidala12, PJD13}).

On the theoretical front, the discussions about quark and gluon
(or parton) contributions to the nucleon spin have led to a better
understanding of fundamental questions related to gauge invariance
of spin-dependent matrix elements of quark and gluon operators,
and the appropriate definitions of parton orbital angular momentum
in QCD (see Ref.~\cite{LL13} and references therein).
Independent of their physical interpretation, extraction of
spin-dependent parton distribution functions (PDFs) is an important
phenomenological pursuit, in which one seeks a consistent description of
data from a variety of experiments within a perturbative QCD framework.  
With the growing body of experimental data being accumulated, and plans
for future experiments at RHIC, COMPASS and Jefferson Lab to achieve
greater precision and access to regions of kinematics hitherto
unexplored, the need exists to develop the theoretical tools necessary
to optimally utilize the new empirical information.

Systematic studies of spin-dependent PDFs have been performed by a
number of collaborations, with next-to-leading order (NLO) analyses
using the standard global fitting methodology undertaken by the
DSSV (de~Florian, Sassot, Stratmann and Vogelsang)
  \cite{DSSV09},
LSS (Leader, Sidorov and Stamenov)
  \cite{LSS10},
BB (Bl\"umlein and B\"ottcher)
  \cite{BB10}, and
AAC (Asymmetry Analysis Collaboration) 
  \cite{AAC09} groups. 
All of these analyses utilize inclusive DIS data on proton, deuteron
and $^3$He targets, while the LSS10 \cite{LSS10} and DSSV09
\cite{DSSV09} PDFs are also constrained by semi-inclusive DIS data.
The DSSV09 group uses in addition the polarized $pp$ scattering data
at NLO, while the AAC \cite{AAC09} fits these data via a $K$-factor
approximation for the NLO corrections.
The more recent NNPDF distributions \cite{NNPDF13} are based
on a neural network approach, while the parametrizations from
Ref.~\cite{BSB02} are inspired by a statistical model.
A new NLO analysis of inclusive and semi-inclusive DIS data
was also recently performed in Ref.~\cite{AKS13}.

Determining the nucleon spin fractions carried by quarks and
gluons requires integrating the spin PDFs over all values of
the momentum fraction $x$.  The small-$x$ region in particular
contributes significantly to the integrals (or lowest moments)
of the distributions, and considerable effort has been made to
understand the spin PDFs in this region both experimentally
and theoretically.  On the other hand, at large values of $x$
the $u$ and $d$ quark PDFs are very sensitive to the dynamics
underlying the breaking of SU(2)$\times$SU(2) spin-flavor symmetry
\cite{MT96, Isgur99, Holt10}, as well as to the presence of quark
orbital angular momentum in the nucleon \cite{Avakian07}.
Unfortunately, PDFs are notoriously difficult to extract at
high values of $x$ due to the rapidly dropping cross sections
in this region, and consequently spin structure measurements
have received relatively little attention for $x \gtrsim 0.5$.

More recently, experiments at Jefferson Lab utilizing the high
lumonisities and polarized beams available with the CEBAF accelerator
have collected high-precision data on polarization asymmetries and
cross sections for both longitudinally and transversely polarized
targets \cite{Chen11}.  The new data provide a unique, if somewhat
limited, window on the high-$x$ region, which will be further
extended following the 12~GeV energy upgrade \cite{12GeV}.
In an effort to maximally exploit the new data, the Jefferson Lab
Angular Momentum (JAM) Collaboration \cite{JAMweb} has embarked
on a program to perform a global NLO analysis of world data,
over a large range of kinematics, including a systematic study
of spin-dependent PDFs in the high-$x$ and low-$Q^2$ region.

From the kinematics of inclusive DIS, at fixed four-momentum transfer
squared $Q^2$, increasing values of $x$ are naturally correlated with
decreasing invariant masses $W$ of the produced hadronic system,
	$W^2 = M^2 + Q^2 (1-x)/x$,
where $M$ is the mass of the nucleon.  To ensure that the leading twist
contribution dominates the scattering process, one must restrict $W$
to be large enough for the final state to be generated mostly by
incoherent fragmentation of partons into hadrons, with $W$ above
the region where coherent resonance structures are visible,
	$W \geq W_{\rm res} \approx (1.8-2.0)$~GeV.
At fixed $Q^2$ this means that $x$ is bounded by
	$x \leq x_{\rm res} = Q^2 / (W_{\rm res}^2 - M^2 + Q^2)$,
so that in practice at $Q^2 = 1$~GeV$^2$ the DIS region is
limited to $x \lesssim 0.25$, and even at $Q^2 = 5$~GeV$^2$
it is constrained to $x \lesssim 0.6$.
Since the experimentally explored $Q^2$ range is more restricted
in polarized DIS than in unpolarized DIS, a significantly larger
portion of the spin-dependent data lie in the small-$Q^2$ region.
Consequently, obtaining direct information on PDFs at large $x$
presents an even more difficult task than for spin-averaged PDFs,
which themselves have sizeable uncertainties for $x \gtrsim 0.7$
\cite{CJ12}.  To compensate for the smaller overall number of data
points available to global spin PDF analyses, one is then typically
forced to make use of data down to $Q^2 \approx 1$~GeV$^2$.

The inclusion of data at large $x$ (or low $W$) and low $Q^2$
presents additional challenges for global PDF studies.
To account for deviations in the low-$Q^2$ data from the logarithmic
$Q^2$ dependence expected from perturbative QCD, corrections from
various nonperturbative effects at finite $Q^2$, such as target
mass and higher twist corrections, must be included.
Several of the existing analyses implement some of these effects,
such as the target mass corrections (TMCs) in Refs.~\cite{LSS10,
BB10, NNPDF13}, and higher twist corrections to the $g_1$ structure
function in the LSS10 \cite{LSS10} and BB10 \cite{BB10} fits.
In the present analysis we incorporate both twist-3 and twist-4
corrections to $g_1$, as well as a twist-3 correction to the $g_2$
structure function.

Moreover, when using data obtained from experiments with polarized
deuterium or $^3$He nuclei, whose use is necessary for the separation
of the $u$ and $d$ quark flavors, nuclear effects must be removed.
While existing analyses typically do account for some nuclear effects
through the method of effective polarizations, at large values of $x$
nuclear Fermi motion plays an increasingly important role, requiring
nuclear smearing corrections to be applied, as well as other possible
corrections associated with nucleon off-shell and non-nucleonic degrees
of freedom \cite{Ethier13}.  The nuclear corrections have been shown in
recent unpolarized PDF analyses \cite{CJ12, CJ10, CJ11, ABKM09, ABM12,
MMSTWW13, JR-new} to be essential for correctly describing DIS data at
large $x$ and low $W$, and are even more important in polarized fits,
where low-$W$ DIS data comprise a substantial fraction of the available
data.  In this work we consistently apply the nuclear smearing
corrections to both the $g_1$ and $g_2$ structure functions for
both deuterium and $^3$He, within the framework of the weak binding
approximation \cite{Ethier13, KM08, KM09}, and examine their impact
on the extracted PDFs.

The strategy adopted in most previous global PDF studies has involved
fitting the $g_1$ structure function or the polarization asymmetry
$A_1$, which are typically extracted from the experimental longitudinal
polarization asymmetry using parametrizations of the unpolarized
structure functions and assumptions about the transverse spin
dependent $g_2$ structure function.  In contrast, the JAM analysis
directly fits the measured longitudinal and transverse asymmetries,
where available, thereby eliminating any potential biases introduced
into the analysis by the use of inconsistent unpolarized structure
function inputs obtained from separate experimental analyses.
The unpolarized PDFs used in the JAM analysis represent a new global
fit to the world's spin-averaged data, performed along the lines of
the spin-dependent fit and using a similar set of assumptions,
kinematic cuts, and theoretical nuclear and finite-$Q^2$ corrections
(for details see Ref.~\cite{JR-new}).

With the focus of this study being primarily on PDFs at intermediate
and large $x$, our strategy will be to first systematically explore
the influence of the various corrections on the $u$ and $d$ quark
distributions, whose determination in principle requires only inclusive
DIS data.  Once the basic fits are established, in subsequent work the
JAM analysis will be extended to study sea quark flavor decomposition
and gluon polarization at small $x$ using semi-inclusive DIS and
polarized hadron--hadron scattering data.

The outline of this paper is as follows.
In Sec.~\ref{sec:formalism} we summarize the basic formulas for the
inclusive DIS cross sections and asymmetries, and the polarized
data sets used in this analysis.
The salient features of the theoretical QCD framework are reviewed in
Sec.~\ref{sec:QCD}, including the choice of PDF parametrization and
the analysis of PDF errors.  Here we also present the reference fit
which will be used as the baseline for the subsequent study of the
various strong-interaction effects discussed in the rest of the paper.
The effects of the nuclear smearing corrections in the deuteron and
$^3$He data on the fitted PDFs are described in Sec.~\ref{sec:nuclear},
and those due to finite-$Q^2$ corrections in Sec.~\ref{sec:finiteQ}.
The results of the full JAM fit are presented in Sec.~\ref{sec:JAMfits},
and parameter values for the leading twist and higher twist
contributions to the structure functions are provided.
The impact of the recent high-precision data from Jefferson Lab at
low $W$ and $Q^2$ is discussed in Sec.~\ref{sec:cuts}, together with
the effects of different kinematic cuts in $W$ and $Q^2$ on the
determination of the leading twist PDFs and higher twist corrections.
Finally, in Sec.~\ref{sec:conc} we recount the findings of the
present study, and outline future plans for the JAM PDF analysis.

%%%%%%%%%%%%%%%%%%%%%%%%%%%%%%%%%%%%%%%%%%%%%%%%%%%%%%%%%%%%%%%%%%%%%%%%%
\section{Polarized deep-inelastic scattering}
\label{sec:formalism}

In this section we briefly review the definitions of the inclusive
DIS cross sections and polarization asymmetries that will provide the
data to be fitted in the JAM analysis.  We utilize data on both the
longitudinal and transverse asymmetries for hydrogen, deuterium and
$^3$He targets, or on the $A_1$ and $A_2$ asymmetries where these
are given instead.

% .......................................................................
\subsection{Cross sections and asymmetries}
\label{ssec:def}

Most of the information that currently exists on the partonic spin
structure of the nucleon has come from experiments involving inclusive
scattering of longitudinally polarized leptons from longitudinally
polarized targets.  By taking differences of cross sections with the
spin of the target parallel and antiparallel to that of the lepton,
one measures primarily the $g_1$ structure function of the nucleon,
which in the parton model is expressed in terms of the spin (or helicity)
dependent PDFs.  Information on the $g_2$ structure function, which
does not have a simple partonic interpretation, can be gathered by
measuring differences of cross sections with the target polarized
transversely to the beam polarization.

The asymmetries for longitudinal $(A_{\|})$ and transverse
$(A_{\perp})$ target polarization are defined in terms of the
differential cross sections $\sigma \equiv d^2\sigma/d\Omega dE'$ by
\begin{eqnarray}
A_{\|}
&=& { \sigma^{\uparrow\Downarrow} - \sigma^{\uparrow\Uparrow}
\over \sigma^{\uparrow\Downarrow} + \sigma^{\uparrow\Uparrow} }\ ,
\ \ \ \ \
A_{\perp}\,
=\, { \sigma^{\uparrow\Rightarrow} - \sigma^{\uparrow\Leftarrow}
\over \sigma^{\uparrow\Rightarrow} + \sigma^{\uparrow\Leftarrow} }\ ,
\label{eq:A_def}
\end{eqnarray}
where the arrows $\uparrow$ and $\Uparrow (\Downarrow)$ denote
the electron and nucleon spins in the same (opposite) directions,
respectively, with the corresponding nucleon polarizations transverse
to the beam direction labeled by $\Rightarrow$ and $\Leftarrow$.
It is convenient also to define the spin asymmetries $A_1$ and $A_2$,
such that
\begin{equation}
A_\parallel\, =\, D (A_1 + \eta A_2),
\hspace{0.7cm}
A_\perp\, =\, d (A_2 - \xi A_1),
\end{equation}
where the kinematical variables are given by
\begin{subequations}
\begin{equation}
D\, =\, \frac{y(2-y)(2+\gamma^2 y)}
	     {2(1+\gamma^2) y^2 + (4(1-y)-\gamma^2 y^2)(1+R)},
							\hspace{0.7cm}
\eta\, =\, \gamma\, \frac{4(1-y)-\gamma^2 y^2}
			 {(2-y)(2+\gamma^2 y)},
\end{equation}
\begin{equation}
d\, =\, \frac{\sqrt{4(1-y)-\gamma^2 y^2}}{2-y}\, D,
							\hspace{0.7cm}
\xi\, =\, \gamma\, \frac{2-y}{2+\gamma^2 y}.
\end{equation}
\end{subequations}%
Here $y = \nu/E = Q^2/2MxE$ is fraction of the incident lepton's
energy $E$ transferred to the target, $\nu$ and $Q^2$ are the energy
transfer and four-momentum transfer squared (virtuality of the photon),
respectively, and $x = Q^2/2M\nu$ is the Bjorken scaling variable,
with $M$ the nucleon mass, and $\gamma^2 = 4 M^2 x^2/Q^2$.
The function $R$ is the ratio of cross sections for longitudinal
to transversely polarized virtual photons,
\begin{equation}
R\, =\, \frac{F_L}{(1+\gamma^2) F_2 - F_L},
\end{equation}
with the longitudinal structure function $F_L$ defined in terms
of the unpolarized $F_1$ and $F_2$ structure functions as
\begin{equation}
F_L\, =\, (1+\gamma^2)F_2 - 2xF_1.
\end{equation}

The $A_1$ and $A_2$ asymmetries are directly related to the spin
structure functions by
\begin{equation}
A_1\ =\ (g_1-\gamma^2 g_2)\frac{2x}{(1+\gamma^2) F_2 - F_L},
\hspace{0.7cm}
A_2\ =\ \gamma (g_1 + g_2)\frac{2x}{(1+\gamma^2) F_2 - F_L}.
\label{eq:A1A2}
\end{equation}
Since $x$ is bounded by $0 \leq x \leq 1$, at high energies the
longitudinally polarized cross section is dominated by the $g_1$
structure function, with $A_1 \to g_1/F_1$ in the high-$Q^2$ limit,
while both $g_1$ and $g_2$ contribute at the same order to the 
transversely polarized cross section.
At finite $Q^2$, however, knowledge of both unpolarized structure
functions is necessary for extracting the $g_1$ and $g_2$ structure
functions, and the spin-dependent parton distributions from the
inclusive asymmetries.

% .......................................................................
\subsection{Data}
\label{ssec:data}

The inclusive DIS data sets considered in this analysis include
polarization asymmetries of the proton, deuteron and $^3$He measured at
CERN in the
  EMC \cite{EMC89},
  SMC \cite{SMC98, SMC99} and
  COMPASS \cite{COMPASS10, COMPASS07} experiments, at
SLAC with the
  E130 \cite{SLAC-E130},
  E142 \cite{SLAC-E142},
  E143 \cite{SLAC-E143},
  E154 \cite{SLAC-E154},
  E155 \cite{SLAC-E155p, SLAC-E155_A2pd, SLAC-E155d} and
  E155x \cite{SLAC-E155x} experiments,
by the
  HERMES Collaboration \cite{HERMES07, HERMES97, HERMES12} at
DESY, as well as more recent experiments in
  Hall~A \cite{E99-117, E97-103} and
  Hall~B \cite{EG1a, EG1b} at
Jefferson~Lab.

\begin{table}[p]
\caption{Inclusive spin-dependent DIS data sets used in the JAM
	analysis, indicating the type of asymmetry measured and
	the number of data points $N_{\rm dat}$ within the cuts,
	as well as the contribution of each data set to the overall
	$\chi^2$ of the fits (for the full JAM and leading twist fits).
	For the HERMES ``$n$'' measurement \cite{HERMES97}, only
	the extracted ``neutron'' $A_1$ asymmetry is available;
	for the E155x transverse asymmetries $\tilde{A}_\perp$
	\cite{SLAC-E155x} the target polarizations were not exactly
	perpendicular to the beam line.  The data sets marked with
	asterisks ($^*$) are not used in the full JAM fits,
	but are discussed in Sec.~\ref{sec:finiteQ}.\\ }
\scriptsize 
\begin{tabular}{lc|ccc|cc}			\hline\hline
experiment
	&\ reference\
	&\ observable\
	&\ target\
	&\ $N_{\rm data}$\
	&\ $\chi^2({\rm LT})/N_{\rm dat}$\		
	&\ $\chi^2({\rm JAM})/N_{\rm dat}$\	\\ \hline
EMC	& \cite{EMC89}
	& $A_1$
	& $p$
	& 10			
	& 0.42					
	& 0.39					\\
SMC	& \cite{SMC98}
	& $A_1$
	& $p$
	& 12
	& 0.36
	& 0.36					\\
	& \cite{SMC98}
	& $A_1$
	& $d$
	& 12
	& 1.59					
	& 1.66					\\
	& \cite{SMC99}
	& $A_1$
	& $p$
	& 8
	& 1.37
	& 1.35					\\
	& \cite{SMC99}
	& $A_1$
	& $d$
	& 8
	& 0.54				
	& 0.56					\\		
COMPASS	& \cite{COMPASS10}
	& $A_1$
	& $p$
	& 15
	& 0.95
	& 0.97					\\
	& \cite{COMPASS07}
	& $A_1$
	& $d$
	& 15
	& 0.57				
	& 0.51					\\
SLAC E80/E130
	& \cite{SLAC-E130}
	& $A_\parallel$
	& $p$
	& 23
	& 0.52				
	& 0.54					\\
SLAC E142
	& \cite{SLAC-E142}
	& $A_1$
	& $^3$He
	& 8	
	& 0.58		
	& 0.70					\\
	& \cite{SLAC-E142}
	& $A_2$
	& $^3$He
	& 8	
	& 0.70						
	& 0.70					\\
SLAC E143
	& \cite{SLAC-E143}
	& $A_\parallel$
	& $p$
	& 85
	& 0.85				
	& 0.81					\\
	& \cite{SLAC-E143}
	& $A_\perp$
	& $p$
	& 48
	& 0.95
	& 0.91					\\
	& \cite{SLAC-E143}
	& $A_\parallel$
	& $d$
	& 85	
	& 1.05					
	& 0.85					\\
	& \cite{SLAC-E143}
	& $A_\perp$
	& $d$
	& 48	
	& 0.92						
	& 0.91					\\
SLAC E154
	& \cite{SLAC-E154}
	& $A_\parallel$ 
	& $^3$He
	& 18	
	& 0.43			
	& 0.42					\\
	& \cite{SLAC-E154}
	& $A_\perp$
	& $^3$He
	& 18
	& 1.00							
	& 1.00					\\
SLAC E155
	& \cite{SLAC-E155p}
	& $A_\parallel$
	& $p$
	& 73	
	& 1.00					
	& 0.92					\\
	& \cite{SLAC-E155p, SLAC-E155_A2pd}
	& $A_\perp$
	& $p$
	& 66	
	& 1.00				
	& 0.96					\\
	& \cite{SLAC-E155d}
	& $A_\parallel$
	& $d$
	& 73	
	& 0.98	
	& 0.97					\\
	& \cite{SLAC-E155d, SLAC-E155_A2pd}
	& $A_\perp$
	& $d$
	& 66	
	& 1.51			
	& 1.49					\\
SLAC E155x
	& \cite{SLAC-E155x}
	& $\tilde{A}_\perp$
	& $p$
	& 117
	& 2.17		
	& 1.64					\\
	& \cite{SLAC-E155x}
	& $\tilde{A}_\perp$
	& $d$
	& 117	
	& 0.90	
	& 0.84					\\
HERMES	& \cite{HERMES07}
	& $A_\parallel$
	& $p$
	& 37
	& 0.38	
	& 0.39					\\
	& \cite{HERMES07}
	& $A_\parallel$
	& $d$
	& 37
	& 0.86							
	& 0.85					\\
	& \cite{HERMES97}
	& $A_1$
	& ``$n$''
	& 9
	& 0.29
	& 0.30					\\
	& \cite{HERMES12}
	& $A_2$
	& $p$
	& 20
	& 1.07	
	& 1.16					\\
JLab E99-117
	& \cite{E99-117}
	& $A_\parallel$
	& $^3$He
	& 3	
	& 0.62		
	& 0.06					\\		
	& \cite{E99-117}
	& $A_\perp$
	& $^3$He
	& 3	
	& 1.08	
	& 0.87					\\
COMPASS	& \cite{COMPASSg}
	& $\Delta g/g$
	& $p$
	& 1
	& 5.27		
	& 2.71					\\ \hline
total	&
	& 
	&
	& 1043
	& 1.07		
	& 0.98					\\ \hline
JLab E97-103$^*$
	& \cite{E97-103}
	& $A_\parallel$
	& $^3$He
	& 2	
	& ---							
	& ---					\\
	& \cite{E97-103}
	& $A_\perp$
	& $^3$He
	& 2	
	& ---							
	& ---					\\
JLab EG1b$^*$
	& \cite{EG1b}
	& $A_1$
	& $p$
	& 766	
	& ---							
	& ---					\\
(prelim.)& \cite{EG1b}
	& $A_1$
	& $d$
	& 767	
	& ---							
	& ---					\\ \hline
\end{tabular}
\label{tab:data}
\end{table}
\normalsize

These data sets are summarized in Table~\ref{tab:data}, which lists
the relevant observables available from each experiment, and the
number of data points that lie within the $Q^2$ and $W^2$ cuts.
(The data sets can also be found in the online JAM Database
\cite{JAMdb}.)
The nominal cuts used in the JAM analysis are
  $Q^2 \geq 1$~GeV$^2$ and
  $W^2 \geq 3.5$~GeV$^2$.
The latter is slightly lower than in some previous spin PDF analyses,
and both are significantly smaller than in many unpolarized global
PDF analyses, which reflects the more limited range of data available
from polarized DIS experiments.  In Sec.~\ref{sec:cuts} we study the
effect on the spin-dependent PDFs of varying these cuts.

For the CERN data \cite{EMC89, SMC98, SMC99, COMPASS10, COMPASS07},
only $A_1$ asymmetries are given; since the data are typically
taken at high energies, the $A_2$ contribution to the measured
$A_\parallel$ asymmetry is small.
The HERMES $p$ and $d$ data are given in terms of the $A_\parallel$
asymmetry \cite{HERMES07}, while the earlier ``neutron'' data
(extracted from $^3$He) are given in terms of $A_1$ \cite{HERMES97}.
The more recent transverse polarization analysis \cite{HERMES12}
measured the $A_2$ asymmetry of the proton.
For most of the SLAC experiments on $p$ and $d$ targets, both
$A_\parallel$ and $A_\perp$ were measured; for $^3$He targets,
the earlier E142 experiment \cite{SLAC-E142} presented results for
$A_1$ and $A_2$, while the later E154 experiment \cite{SLAC-E154}
provided $A_\parallel$ and $A_\perp$.
For the E155x experiment on $p$ and $d$, the target polarizations
were not exactly perpendicular to the beam line, but at an angle,
and in this analysis we use the exact kinematics as given in
Ref.~\cite{SLAC-E155x}.
Finally, experiments in Hall~A at Jefferson Lab with $^3$He
targets obtained the $A_\parallel$ and $A_\perp$ asymmetries
\cite{E99-117, E97-103}, while the Hall~B data \cite{EG1a, EG1b}
on $p$ and $d$ targets were given in terms of $A_1$.
Note that the data from the new EG1b experiment in Hall~B \cite{EG1b}
supercede the earlier results from the EG1a experiment \cite{EG1a}.
However, at present the EG1b data have not yet been fully analyzed
to enable them to be used in the full JAM fit, although we will
examine the possible effects of the preliminary results on the
global fits in Sec.~\ref{sec:cuts}.

The inclusive DIS data are supplemented with a model-dependent
extraction of the gluon $\Delta g/g$ ratio from semi-inclusive DIS
at COMPASS \cite{COMPASSg}.  While this is not ideal, in practice it
may be reasonable to include these data since the gluon contribution
to the polarization asymmetries only contributes at subleading order
(and through QCD evolution) to inclusive DIS.
A more detailed analysis of the polarized gluon sea requires data
from polarized $pp$ scattering, which will be addressed in a future
study \cite{JAM-III}.

Also listed in Table~\ref{tab:data} are the $\chi^2$ values
for each of the data sets, as well as the total, for the main
JAM fit, as well as from a fit which includes leading twist (LT)
contributions only (see Sec.~\ref{sec:finiteQ}).
Generally, the $\chi^2$ values are smaller for the JAM fit,
which includes higher twist (HT) and other hadronic and nuclear
corrections, than for the fit with leading twist contributions
only.  The improvement in the overall $\chi^2/N_{\rm dat}$ with
the higher twist corrections is in fact quite significant,
indicating a clear preference of the data for the presence
of higher twist effects.

%%%%%%%%%%%%%%%%%%%%%%%%%%%%%%%%%%%%%%%%%%%%%%%%%%%%%%%%%%%%%%%%%%%%%%%%%
\section{QCD framework}
\label{sec:QCD}

In this section we outline the theoretical framework used in the JAM
global QCD analysis of spin-dependent PDFs.  We begin by summarizing
the pertinent results for the structure functions in terms of leading
twist PDFs at NLO, before discussing our choice of parametric forms
for the individual parton distributions and their constraints. 
We also discuss the treatment of PDF errors, and present a reference
fit which will be used as a baseline to study the impact of various
hadronic and nuclear corrections in subsequent sections.

% .......................................................................
\subsection{Structure functions at leading twist}
\label{ssec:LT}

In the leading twist approximation, the factorization theorems of
QCD allow the $g_1$ structure function to be expressed in terms of
spin-dependent (or helicity) quark and gluon distribution functions.
For convenience, we work in moment space, where the $n$-th Mellin moment
of a PDF $\Delta f(x,Q^2)$ ($f=q$, $\bar q$ or $g$) is defined as   
\begin{equation}
\Delta f^{(n)}(Q^2) = \int_0^1 dx \; x^{n-1} \Delta f(x,Q^2).
\end{equation}
In massless leading twist QCD, the $n$-th moments of the $g_1$
structure function can be written as
\begin{eqnarray}
g_1^{(n)}(Q^2)
&=& \frac{1}{2} \sum_q e_q^2\;
    \left( \Delta C_{qq}^{(n)}(Q^2)\, \Delta q^{(n)}(Q^2)
         + \Delta C_g^{(n)}(Q^2)\, \Delta g^{(n)}(Q^2)
    \right),
\label{eq:g1LT}
\end{eqnarray}
where the moments of the quark and gluon hard scattering coefficient
functions $\Delta C_{qq}^{(n)}$ and $\Delta C_g^{(n)}$ are calculable
perturbatively, and are summarized in Refs.~\cite{Lampe98, Weigl96}
up to NLO.

In the cross sections, or $A_1$ and $A_2$ asymmetries in
Eq.~(\ref{eq:A1A2}), the $g_2$ structure function is always
suppressed by a power of $\gamma \sim M/Q$.  Strictly speaking,
therefore, in the massless \mbox{($Q^2 \to \infty$)} limit only
the $g_1$ structure function contributes.  If one considers also
$g_2$, for consistency one needs to include also target mass
corrections to $g_1$, as we discuss in Sec.~\ref{sec:finiteQ}.
Furthermore, operators containing masses mix with higher twist
operators under renormalization, so that in practice the $g_2$
structure function contains twist $\tau=3$ contributions in addition
to $\tau=2$.  The latter is given through the Wandzura-Wilczek
relation in terms of the $\tau=2$ contribution to the $g_1$
structure function \cite{Wandzura77},
\begin{eqnarray}
g_2^{(n)}(Q^2)_{\tau=2}
&=& -\frac{n-1}{n} g_1^{(n)}(Q^2)_{\tau=2}.
\label{eq:g2WW}   
\end{eqnarray}
In Sec.~\ref{sec:finiteQ} we will also consider higher twist
contributions to $g_2$, in addition to $g_1$.
While the $\tau=2$ part of the lowest ($n=1$) moment of the $g_2$
structure function obviously satisfies the Burkhardt-Cottingham (BC)
sum rule \cite{BC70},
\begin{eqnarray}
g_2^{(1)}(Q^2) &=& 0,
\label{eq:g2BC}
\end{eqnarray}
whether this is also satisfied by the higher twist contributions
will be discussed in Sec.~\ref{sec:finiteQ}.
Note that the leading polarized quark coefficient function
  $\Delta C_{qq}^{(n)} = 1 + {\cal O}(\alpha_s)$
contributes already at LO, while the polarized gluon enters only at NLO,
  $\Delta C_{qg}^{(n)} = {\cal O}(\alpha_s)$.

% .......................................................................
\subsection{Parton distributions}
\label{ssec:param}

The scale dependence, or evolution, of the polarized PDFs has been
calculated up to NLO in perturbative QCD, and has the same structure
as in the unpolarized case.  We follow closely the formalism adopted
in Ref.~\cite{JR09} for the evolution of the Mellin moments of the
distributions.  In particular, we use the so-called {\em truncated}
solutions, in which subleading terms are explicitly removed from
the solution of the evolution equations.  The splitting functions
appropriate for the polarized NLO evolution can be found, for example,
in Ref.~\cite{Lampe98}.

Since the PDF scale dependence is completely specified by the QCD 
evolution equations, all parton distributions at any scale are
specified by the values of a complete set of distributions at
the input scale, for which we have chosen $Q_0^2=1$~GeV$^2$.
The input distributions at this scale are parametrized as
\begin{equation}      
x \Delta f(x,Q_0^2)
= N_f\, x^{a_f} (1-x)^{b_f} (1 + c_f \sqrt{x} + d_f x).
\label{eq:param}
\end{equation}
for $f = u^+, d^+, \bar{u}, \bar{d}, \bar{s}, g$,
where $q^+ = q +\bar{q}$ and we assume $\Delta\bar{s}=\Delta s$.
The choice of basis functions $\Delta u^+$ and $\Delta d^+$,
rather than, say, the valence
  $\Delta u-\Delta \bar{u}$
and
  $\Delta d-\Delta \bar{d}$
distributions, is motivated by the fact that these are the
functions which naturally enter into the $g_1$ structure
function in inclusive DIS, see Eq.~(\ref{eq:g1LT}).

In practice, the parametrization (\ref{eq:param}) is too general
for the information available in our analyses, and additional
constraints have to be adopted.
First, we note that inclusive DIS is only sensitive to three of 
the quark distributions which, for example, can be taken to be
$\Delta u^+$, $\Delta d^+$, and $\Delta s^+$.  The first moments
$\Delta q^{+(1)}$ of these distributions are related to matrix
elements of weak baryon decays through the relations
\begin{subequations}
\label{eq:baryon}
\begin{eqnarray}
\Delta u^{+(1)} - \Delta d^{+(1)}
	&= & 1.269 \pm 0.003,				\\
\Delta u^{+(1)} + \Delta d^{+(1)} - 2 \Delta s^{+(1)}
	&=& 0.586 \pm 0.031,
\end{eqnarray}%
\end{subequations}%
These constraints are implemented as additional ``data points'',
so that the fit receives a $\chi^2$ penalty if it deviates appreciably
from the central values.  In practice, in our fits, these latter are
very well reproduced with $\chi^2$ values close to zero.

Note also that while the distributions are fitted to only \emph{two}
independent observables, $g_1^p$ and $g_1^n$ (with the neutron
extracted from experiments with deuteron or $^3$He targets),
it is in principle possible to determine \emph{three} quark PDFs
from these because of their different $Q^2$ evolution \cite{LSS98}.
However, this requires data of sufficient accuracy at different scales,
and in practice the constraints obtained from the evolution are not
very robust.  For example, in fits with a range of (fixed) values
for $\Delta s^+$, the remaining distributions compensate so that
very similar descriptions of the data are achieved in each case.
To avoid superfluous parameters (flat directions in $\chi^2$) and
overfitting, we leave only $N_{\bar{s}}$ as a free parameter for
$\Delta \bar{s}$, and fix the remaining parameters as
  $a_{\bar{s}} = a_{d^+}$,
  $b_{\bar{s}} = b_g + 2$,
  $c_{\bar{s}} = d_{\bar{s}} = 0$,
and using spectator counting rules at large $x$
\cite{Blankenbecler:1974tm, Farrar:1975yb, BBS95}.

Since $\Delta\bar{u}$ and $\Delta\bar{d}$ do not contribute directly
to the description of the data in our analysis, these distributions
cannot be determined in our fits, and have been fixed by requiring
\begin{equation}
\label{eq:seafix}
\lim_{x\to 0} \Delta \bar{q}(x,Q_0^2)
= \frac{1}{2} \lim_{x\to 0} \Delta q^+(x,Q_0^2),
\end{equation}
for $q=u$ and $d$, which implies
  $N_{\bar{q}} = N_{q^+}/2$ and
  $a_{\bar{q}} = a_{q^+}$.
In addition, we choose
  $b_{\bar{q}} = b_g + 2$ and
  $c_{\bar{q}} = d_{\bar{q}} = 0$.
For our nominal results we have refrained from considering a
symmetric sea ($\Delta \bar{u} = \Delta \bar{d} = \Delta \bar{s}$),
as has been assumed in some previous analyses \cite{AAC09, BB10},
since fits that utilize semi-inclusive DIS data \cite{DSSV09, LSS10}
typically find a non-symmetric sea --- see Fig.~\ref{fig:ref}.
(On the other hand, we found that the sea-symmetric assumption
provides a comparable description of the data.)
In fact, due to the resulting rather flexible parametrization of 
$\Delta u^+$ and $\Delta d^+$, the overall size of $\Delta \bar{u}$
and $\Delta \bar{d}$ is poorly constrained.  To avoid unphysical
results and provide reasonable values for all distributions,
we impose in addition the constraints
\begin{equation}
\frac{1}{2}
\left(\left| \frac{\Delta \bar{q}^{(2)}}{\Delta \bar{s}^{(2)}} \right|
    + \left| \frac{\Delta \bar{s}^{(2)}}{\Delta \bar{q}^{(2)}} \right|
\right) = 1 \pm 0.25.
\end{equation}
Namely, we enforce that the integrals of $x\Delta \bar{u}$ and
$x\Delta \bar{d}$ are comparable to that of $x\Delta \bar{s}$
within a factor of $\approx 2$, although no constraint is imposed
on the relative sign of their difference (note that our parametrizations
do not allow for nodes in any of these distributions).
We stress, however, that the polarized antiquark distributions in the
JAM analysis are not fitted directly, but rather determined by the
specific choices of parameters outlined above.  They are considered
mostly because of the need for completeness of the set of $Q^2$
evolution equations, and should clearly not be viewed as predictions.

As noted in Sec.~\ref{ssec:LT}, the gluon distribution $\Delta g$
only contributes directly to the polarized structure functions at
a subleading level.  In addition, it has some influence on the other
distributions through the QCD evolution.  The data considered in our
analysis provide only mild constraints on this distribution and
in fact a good description of most data can be achieved with
$\Delta g = 0$.
Nevertheless, we have allowed a considerable amount of freedom in
our fits, including nominally $N_g$ and $d_g$ as free parameters,
with the remaining parameters fixed as
$a_g = \frac{1}{2}(a_{u^+} + a_{d^+}) + 1$,
$b_g = \frac{1}{2}(a_{u^+} + a_{d^+}) + 2$,
$c_g = 0$, in order to obtain a reasonable shape for the gluon
distribution.

% .......................................................................
\subsection{Statistical estimation and error analysis}
\label{sec:errors}

The free parameters of the input distributions have been
determined using the formalism detailed in the Appendix of
Ref.~\cite{JimenezDelgado:2012zx}.
This includes a least-squares estimator which takes into account
the correlated systematic uncertainties via analytically
determined nuisance parameters, and an iterative procedure for
an appropriate treatment of multiplicative correlated errors.
Although most of the data sets included in our analysis do not
provide correlated systematic uncertainties, they do usually
include normalizations uncertainties.  A proper treatment of
these is important in order to avoid different biases which
might otherwise occur with more naive treatments
(see Ref.~\cite{JimenezDelgado:2012zx} for details).

The evaluation of our PDF uncertainties is based on the Hessian
method \cite{Pumplin:2001ct}.  We have not observed significant
tensions between different data sets or encountered particularly
flat directions in the parameter space, so that all the free
parameters defined in Sec.~\ref{ssec:param} are included in the
error calculation.
The reported PDF errors in this work refer to variations of
$\Delta\chi^2 = 1$ around the minimum.  Different choices have
also been made in the literature, such as $\Delta\chi^2 = 12.65$
in the AAC analysis \cite{AAC09}, while the DSSV group \cite{DSSV09}
considered both $\Delta\chi^2 = 1$ and $\Delta\chi^2/\chi^2 = 2\%$,
and some unpolarized PDF analyses have used even larger values.
There is no unique criterion for selecting the correct $\chi^2$
interval, and various arguments have been made in favor of
different ways to illustrate the effective uncertainty range
--- see Ref.~\cite{AAC09} for a discussion.
In the JAM analysis we choose the traditional $\Delta\chi^2 = 1$
interval, based on statistical considerations alone.

% .......................................................................
\subsection{Reference fit}
\label{ssec:reffit}

For our baseline reference fit, we include the same data sets as used
in the full JAM fit (listed in Table~\ref{tab:data}), with the same
cuts in $Q^2$ and $W^2$.  However, the reference fit uses only leading
twist contributions, with no target mass or higher twist corrections,
and no nuclear smearing effects.
In particular, this means that the $g_2$ structure function is
approximated by its twist-2 (Wandzura-Wilczek) contribution,
Eq.~(\ref{eq:g2WW}).
As in the full JAM analysis, we fit proton, deuteron and $^3$He data
directly, rather than the model-dependent experimental extractions
of the neutron.  For the unpolarized PDF fit, we employ the
parametrizations from Ref.~\cite{JR-new}, which are obtained under
similar set of assumptions, kinematic cuts, and theoretical inputs
as the JAM fit.
The reference fit so constructed allows us to clearly identify
the various effects that are introduced in the full JAM analysis.

\begin{figure}[p]
\includegraphics[width=15cm]{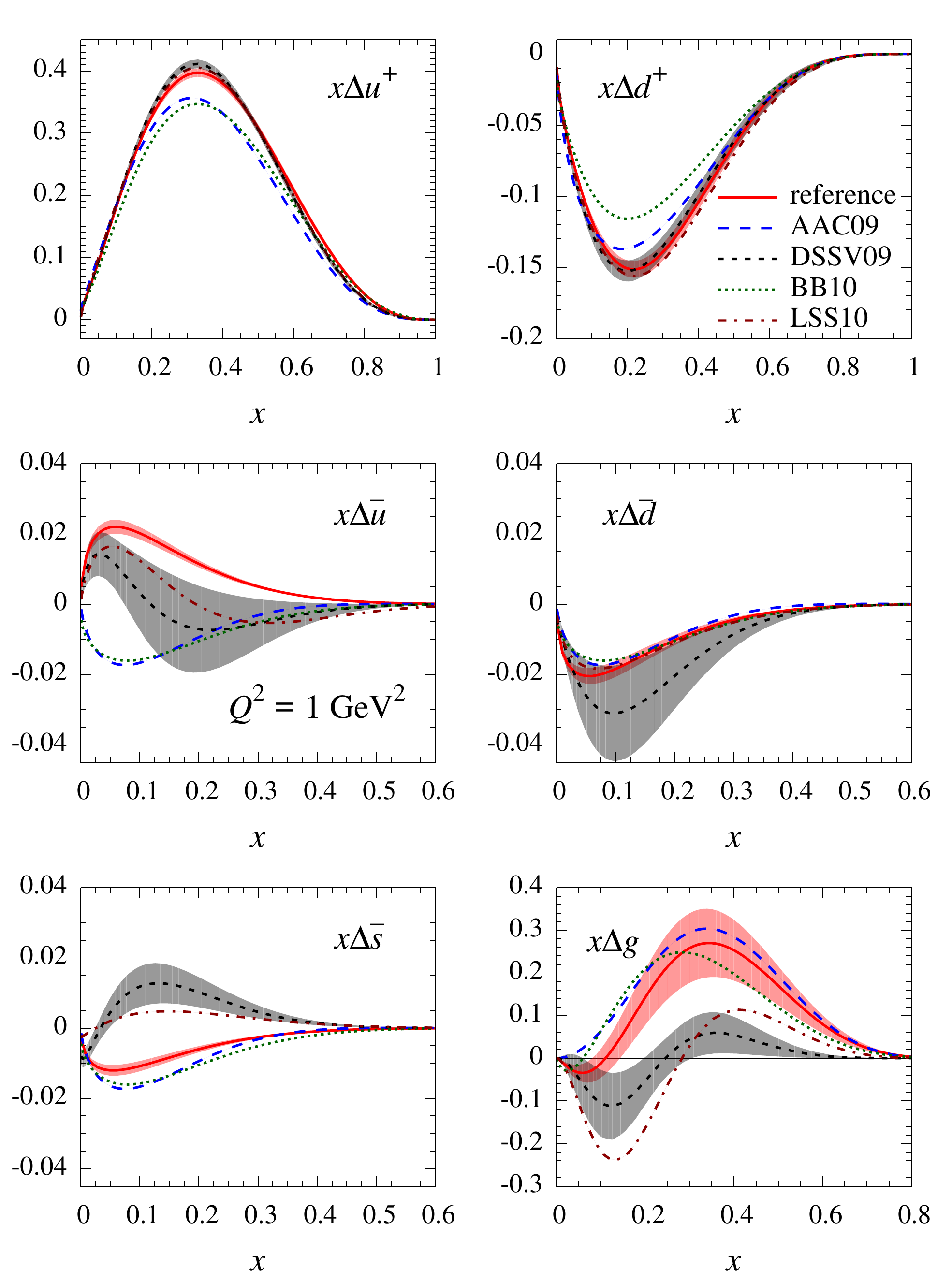}
\caption{Spin-dependent parton distributions for the
	$\Delta u^+$, $\Delta d^+$, $\Delta \bar u$, $\Delta \bar d$
	and $\Delta \bar s$ quark flavors and the polarized gluon
	$\Delta g$ at a scale of $Q^2 = 1$~GeV$^2$.
	The reference JAM fit (red solid) is compared with the recent
	AAC09 \cite{AAC09} (blue dashed),
	DSSV09 \cite{DSSV09} (black short-dashed),
	BB10 \cite{BB10} (green dotted) and
	LSS10 \cite{LSS10} (brown dot-dashed)
	parametrizations.}
\label{fig:ref}
\end{figure}

In Fig.~\ref{fig:ref} we compare the results of the reference JAM fit
for the polarized $\Delta u^+$, $\Delta d^+$, $\Delta \bar u$,
$\Delta \bar d$ and $\Delta \bar s$ quark and polarized gluon
distributions with several recent parametrizations
\cite{DSSV09, LSS10, BB10, AAC09} at a scale of $Q^2 = 1$~GeV$^2$.
For clarity, we show the uncertainty bands for the reference and DSSV09
PDFs \cite{DSSV09} only, and for the others just the central values.
The $\Delta u^+$ PDF is the best constrained polarized distribution,
over a relatively broad range of $x$, due mostly to the proton structure
function data.  The corresponding $\Delta d^+$ distribution has somewhat
larger uncertainties, on the other hand, especially at high $x$ values,
since it requires the relatively more scarce $^3$He (and to a lesser
extent deuteron) data.
The reference JAM $\Delta u^+$ and $\Delta d^+$ PDFs turn out to be very
similar to the DSSV09 and LSS10 distributions, and slightly larger in
magnitude at intermediate $x$ compared with the AAC09 and BB10 results.
The uncertainty bands for $\Delta u^+$ and $\Delta d^+$ are smaller
than the total variation between the different parametrizations,
which is likely related to the systematic uncertainties associated
with the data set choices and theoretical inputs (such as nuclear
and finite-$Q^2$ corrections) being larger than the experimental
uncertainties.

On the other hand, the sea quark polarization is considerably smaller
than the total (or valence), with significantly larger errors.
Note that while in principle some constraint on the polarized strange
distribution $\Delta \bar s$ could be obtained from the inclusive DIS
data through its $Q^2$ evolution, in practice this is challenging
because of the limited precision of the $g_1$ data.
As a result the polarized sea quark distributions are more strongly
dependent upon assumptions about flavor symmetry of the proton sea.
At present even the semi-inclusive DIS data do not provide conclusive
evidence of a nonzero light quark sea, although a slight trend is
indicated towards a more negative $\Delta \bar d$ distribution and a
positive $\Delta \bar u$.
In particular, while the $\Delta \bar d$ distribution is negative
for all the parametrizations shown in Fig.~\ref{fig:ref}, the sign
of the $\Delta \bar u$ and $\Delta \bar s$ PDFs differs for the
different fits at intermediate $x$ values.  Namely, while the
AAC09 and BB10 PDFs, which do not utilize semi-inclusive DIS data,
have negative polarized sea distributions, the DSSV09 and LSS10
$\Delta \bar s$ parametrizations are positive for $x \gtrsim 0.05$,
but negative at smaller $x$.  The $\Delta \bar u$ distribution is
positive at $x \lesssim 0.1-0.2$ for both the DSSV09 and LSS10 fits,
but changes signs at larger $x$ values.
For the reference JAM fit, the polarized strange PDF is chosen
to be negative over all $x$, while $\Delta \bar u$ is positive.
The signs of the $\Delta \bar u$ and $\Delta \bar d$ distributions
here are in fact determined by those of $\Delta u^+$ and $\Delta d^+$,
respectively, with only the sign of $\Delta \bar s$ left as a free
parameter.
The error bands on the JAM antiquark PDFs in Fig.~\ref{fig:ref}
arise from the residual uncertainties on the parameters for the
$\Delta q^+$ distributions, which are related in our analysis to the
antiquark PDF parameters [see Eq.~(\ref{eq:seafix}), for example].
These are not the total uncertainties on the antiquark PDFs,
which would need to be determined from a fit to semi-inclusive
DIS data.
Future measurements of neutral and charged current DIS at an
Electron-Ion Collider (EIC) \cite{Aschenauer:2012ve, Aschenauer:2013iia,
Accardi:2012qut, Accardi:2011mz} will help to clarify the behavior of
the polarized sea quarks.

For the polarized gluon distribution, the indirect constraints
from the $Q^2$ dependence of the inclusive DIS data suggest
a positive $\Delta g$, albeit with very large errors, within
which the data are also compatible with $\Delta g = 0$.
This is in contrast with the COMPASS NLO determination of
$\Delta g$ from open charm muonproduction, which finds
  $\Delta g/g = -0.13 \pm 0.15 \pm 0.15$
at an average
  $\langle Q^2 \rangle \approx 13$~GeV$^2$ and
  $\langle x \rangle \approx 0.2$ \cite{COMPASSg}.
This tension results in the larger $\chi^2$ values for the
COMPASS $\Delta g/g$ point than for all other data sets.
On the other hand, the COMPASS extraction of $\Delta g/g$ follows
a very specific, model-dependent strategy for the extraction.
It will be interesting to explore the consequences of this when
data from polarized $pp$ scattering are included in our subsequent
analysis \cite{JAM-III}.

The reference JAM $\Delta g$ is similar in shape to the AAC09 and
BB10 fit results, which have a slightly more positive distribution
at intermediate $x$ than the DSSV09 and LSS10, which are also more
negative at smaller $x$, $x \sim 0.1$.  We note, however, that our
polarized gluon is essentially unconstrained except for the
normalization and the coefficient of the $(1+x)$ term.
Currently most of the information on $\Delta g$ comes from charm
production in semi-inclusive DIS at COMPASS \cite{Adolph:2012ca},
and from polarized $pp$ scattering with inclusive pion and jet
production at RHIC \cite{Adare:2008aa, Adamczyk:2012qj}.
The data are generally consistent with a small value of $\Delta g/g$,
consistent with zero, although new measurements from RHIC \cite{PJD13},
and possibly a future EIC \cite{Aschenauer:2012ve, Accardi:2012qut,
Accardi:2011mz, NNPDFeic}, have the promise of resolving a small
nonzero distribution.

Overall, we obtain a satisfactory fit to the inclusive DIS data with
the reference JAM parametrization.  This fit will allow us to explore
in detail the effects of the various nuclear and hadronic corrections
that will be discussed in the following sections.  Since the existing
inclusive DIS data do not significantly constrain the sea quark and
gluon distributions which are dominant at small $x$, we will focus our
attention on the polarized valence distributions (or rather the total
$\Delta u^+$ and $\Delta d^+$ PDFs) at intermediate and high values
of $x$.

%%%%%%%%%%%%%%%%%%%%%%%%%%%%%%%%%%%%%%%%%%%%%%%%%%%%%%%%%%%%%%%%%%%%%%%%%
\section{Nuclear corrections}
\label{sec:nuclear}

The absence of stable free neutron targets in scattering experiments
has required information about the spin structure of the neutron to
be obtained using polarized light nuclei, such deuterium or $^3$He,
as effective polarized neutron targets.  Since the binding energies
of these nuclei are small compared with the typical momentum transfers
$Q^2$, historically the effects of nuclear binding and Fermi motion
have been assumed to be negligible.
In the static limit, the nuclear effects on the structure functions can
be introduced through effective proton $P_{p/A}$ and neutron $P_{n/A}$
polarizations and in the nucleus $A$, with the nuclear $g_i^A$ ($i=1,2$)
structure functions given by
\begin{eqnarray}
g_i^A(x,Q^2)
&=& P_{p/A}\ g_i^p(x,Q^2) + P_{n/A}\ g_i^n(x,Q^2).
\label{eq:EPA}
\end{eqnarray}
In this effective polarization approximation (EPA) the nuclear effects
are therefore assumed to be independent of $x$.

For the case of the deuteron, the two nucleons can exist in either
an $S$-state, with relative orbital angular momentum $L=0$, or in
a $D$-state, with $L=2$, and share the deuteron spin equally.
For most purposes the relativistic $P$-state contributions,
with $L=1$, which are associated with negative energy contributions,
are negligible \cite{MST94}.
The average polarization of the nucleon $N$ ($p$ or $n$) is then
given by
\begin{eqnarray}
P_{N/d} &=& 1 - {3 \over 2}\, \omega_D,
\end{eqnarray}
where $\omega_D$ is the the deuteron's $D$-state probability.
For realistic models of the deuteron wave function, based on precision
fits to $NN$ scattering data, one finds $\omega_D \approx 5\%-7\%$
\cite{KM08}.

For polarized $^3$He nuclei, most of the time the two protons that
accompany the neutron are paired with opposing spins, so that the
polarization of $^3$He is determined mainly by the neutron.
The nucleons can be in one of several states, most notably
the symmetric $S$-state with a probability $p_S \approx 90\%$,
an $L=2$ $D$-state with probability $p_D \approx 10\%$, and
a mixed-symmetric $S'$-state with a smaller probability,
$p_{S'} \approx 1-2\%$.  In terms of these probabilities, the
average proton and neutron polarizations in $^3$He are given by
\begin{subequations}
\label{eq:Phe3}
\begin{eqnarray}
P_{p/^3{\rm He}}
&=& - {4 \over 3}\, \left( p_D - p_{S'} \right),	\\
P_{n/^3{\rm He}}
&=& p_S - {1 \over 3}\, \left( p_D - p_{S'} \right).
\end{eqnarray}
\end{subequations}%
For realistic $^3$He wave functions computed either by solving
the Faddeev equations for the three-body bound state or by using
variational methods, the dominant neutron polarization is
$\approx 86\%-89\%$, with the proton contributing
$\approx -4\%$ to $-6\%$ \cite{Ethier13, KM09}.
Note that here $P_{p/^3{\rm He}}$ is the {\it total} proton
polarization, rather than the average of the two protons in
$^3$He.  Higher order corrections to Eqs.~(\ref{eq:Phe3}) from
$\bm{p}^2$-weighted moments of the nuclear spectral function,
where $\bm{p}$ is the bound nucleon three-momentum in the nucleus,
tend to reduce the magnitude of the neutron polarization by
$\approx 1\%-1.5\%$, and the proton polarization by
$\approx 2\%-3\%$ of these values \cite{Ethier13}.

While the EPA may be a reasonable approximation at low and intermediate
values of $x$, at large $x$ where nuclear smearing begins to play a
more important role one expects this to break down.  In the large-$x$
region the effects of Fermi motion and nuclear binding can be
incorporated through convolutions of longitudinally polarized nucleon
light-cone distribution functions $f_{ij}^{N/A}$ and bound nucleon
structure functions $g^N_{1,2}$,
\begin{eqnarray}
g_i^A(x,Q^2)
&=& \sum_{N=p,n} f^{N/A}_{ij}(x,\gamma) \otimes g_j^N(x,Q^2),
\hspace*{1cm}  i,j = 1,2
\label{eq:g1conv}
\end{eqnarray}
where the convolution $\otimes$ is defined as
$(f \otimes g)(x) = \int (dz/z) f(z) g(x/z)$,
and the momentum distribution functions $f^{N/A}_{ij}$ in general
depend on the light-cone momentum fraction $z$ of the nucleus
carried by the active nucleon, as well as on the finite-$Q^2$
parameter $\gamma$.
Explicit forms for the spin-dependent light-cone distributions
(or ``smearing functions'') $f^{N/A}_{ij}$ have been computed
by a number of authors for polarized
  deuterons \cite{FS83, Kaptari94, MPT95, KMPW, Ciofi96}
and
  $^3$He \cite{Ciofi93, SS93, Bissey01}.
Here we shall utilize the smearing functions computed within
the weak binding approximation (WBA) by Kulagin {\it et al.}
for the deuteron \cite{KM08} and $^3$He targets \cite{KM09}.
Additional corrections to Eq.~(\ref{eq:g1conv}) from the possible
modification of the nucleon structure functions in the nuclear
medium, as well as from non-nucleonic components of the nuclear
wave function, have been considered \cite{Ethier13, BGST02},
but are generally expected to be small on the scale of the
current experimental uncertainties.

\begin{figure}[t]
\includegraphics[width=10cm]{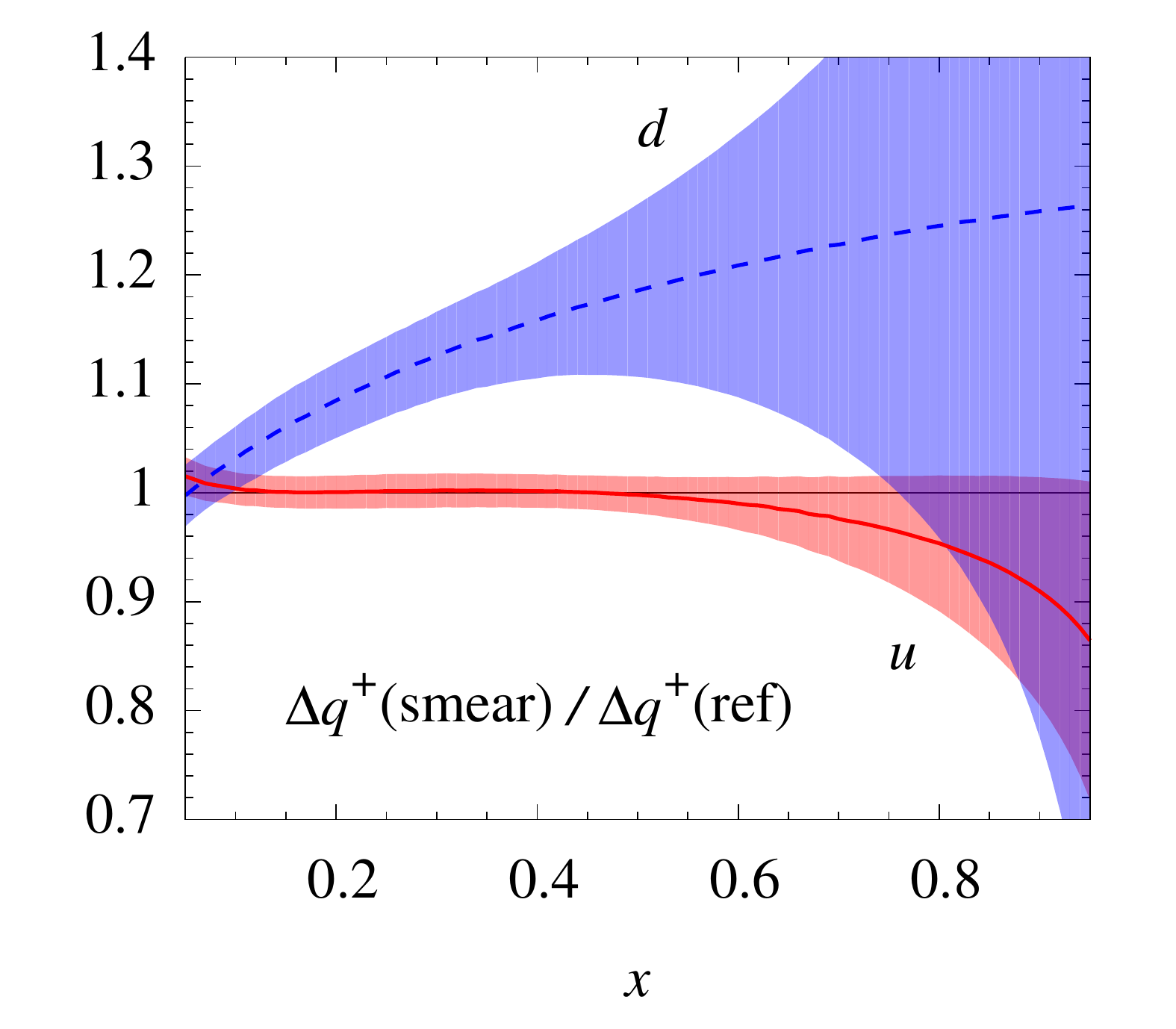}
\caption{Ratios of spin-dependent $\Delta u^+$ and $\Delta d^+$
	distributions, fitted using nuclear smearing corrections,
	to the reference distributions which use the effective
	polarization approximation, at a scale $Q^2 = 1$~GeV$^2$.}
\label{fig:nuclear}
\end{figure}

The effects of the nuclear smearing corrections on the $\Delta u^+$
and $\Delta d^+$ PDFs are illustrated in Fig.~\ref{fig:nuclear},
where the ratios of the distributions computed using the smearing
functions in Eq.~(\ref{eq:g1conv}) to those approximated by the
effective polarization {\it ansatz} in Eq.~(\ref{eq:EPA}) are shown.
The $\Delta u^+$ quark distribution is almost completely inert
to the nuclear smearing models over most of the $x$ range,
$x \lesssim 0.7$, which reflects the fact that it is determined
mainly by the proton data.  A $10\%-20\%$ suppression is seen
at higher $x$ values, although, here the PDFs are essentially
unconstrained by data.

A significantly greater impact of the nuclear smearing corrections
is visible for the $\Delta d^+$ distribution, which is increasingly
enhanced for larger values of $x$, reaching $\approx 20\%-30\%$
at $x \gtrsim 0.5$ for the central values. 
In this region the $\Delta d^+$ distribution is constrained mostly
by the polarized $^3$He data, which at LO is primarily determined
by the combination $\Delta u^+ + 4 \Delta d^+$.
The effect of the nuclear smearing correction is to decrease the
magnitude of the $^3$He polarization asymmetry (or $g_1$ structure
function) at $x \gtrsim 0.5$ \cite{Ethier13}, resulting in a
compensating increase in the magnitude of the (negative)
$\Delta d^+$ distribution in this region.
Including the PDF uncertainties, the enhancement can be even
larger, ranging from $10\%-25\%$ at $x = 0.5$ and $\gtrsim 50\%$
for $x > 0.75$.
Nuclear smearing corrections are thus vital to take into account
if one is to accurately determine the $\Delta d^+$ distribution
at large values of $x$.
Similar conclusions have also been reached in connection with the
extraction of the unpolarized $d$ quark PDF from deuterium data
at high $x$ \cite{CJ12, CJ10}.

%%%%%%%%%%%%%%%%%%%%%%%%%%%%%%%%%%%%%%%%%%%%%%%%%%%%%%%%%%%%%%%%%%%%%%%%%
\section{Finite-$Q^2$ corrections}
\label{sec:finiteQ}

The standard global PDF fitting machinery is constructed in order to
analyze data in the high-$Q^2$ limit, where $Q^2$ and $W^2$ are both
$\gg M^2$, away from the region where nucleon resonances or subleading
effects in $Q^2$ play any significant role.
However, if one were to apply the same $Q^2$ and $W^2$ cuts as are
typically employed in unpolarized global PDF analyses \cite{Forte13},
much of the spin-dependent DIS data would be excluded.
Thus the practical limitations of the polarized DIS data forces us to
utilize low-$Q^2$ and low-$W^2$ data in order to obtain statistically
meaningful fits.
In this section we discuss in detail various finite-$Q^2$ corrections
to the formulas for spin-dependent cross sections or structure
functions in terms of leading twist PDFs, and examine their impact
on the fitted polarized distributions.

In the operator product expansion of QCD, in addition to the usual
twist-2 operators \mbox{$\sim \bar\psi \gamma^\mu \gamma_5\, \psi$}
whose matrix elements give moments of spin-dependent PDFs,
there exist operators involving derivatives, such as
$\bar\psi \gamma^\mu \gamma_5\,
	  D^{\gamma_{\mu_1}} \cdots D^{\gamma_{\mu_n}} \psi$,
which are formally also of twist-2, but enter with additional
powers of $M^2/Q^2$.  Summing these contributions to all orders
in $M^2/Q^2$, one obtains expressions for structure functions
at finite $Q^2$ in terms of massless limit ($M^2/Q^2 \to 0$)
structure functions and their integrals \cite{Matsuda:1979ad,
Piccione:1997zh, Blumlein:1998nv}.
For the moments of the $g_1$ structure function, for instance,
one finds \cite{Blumlein:1998nv}
\begin{eqnarray}
g_1^{(n)}(Q^2)
&=& n \sum_{j=0}^\infty
      \left( \frac{M^2}{Q^2} \right)^j\!
      \frac{(n+j)!}{j!(n-1)!(n+2j)^2}\
      g_{1(0)}^{(n+2j)}(Q^2),
%
% &=& g_1^{(n), 0}(Q^2)\
%  +\ \left( \frac{M^2}{Q^2} \right)
%     \frac{n^2 (n+1)}{(n+2)^2}\, g_1^{(n+2), 0}(Q^2) \nonumber\\
% & &
%  +\ \left( \frac{M^2}{Q^2} \right)^2
%     \frac{n^2 (n+1)(n+2)}{2(n+4)^2}\, g_1^{(n+4), 0}(Q^2) \nonumber\\
% & &
%  +\ \left( \frac{M^2}{Q^2} \right)^4
%     \frac{n^2 (n+1)(n+2)(n+3)}{6(n+6)^2}\, g_1^{(n+6), 0}(Q^2)\
%  +\ {\cal O}\left( \frac{M^8}{Q^8} \right),
%
\end{eqnarray}
where
  $g_{1(0)}^{(n+2j)}(Q^2) = \lim_{M \to 0}\, g_1^{(n+2j)}(Q^2)$
is the $(n+2j)$-th moment of the leading twist structure function in
the $M^2/Q^2 \to 0$ limit.

\begin{figure}[t]
\includegraphics[width=8cm]{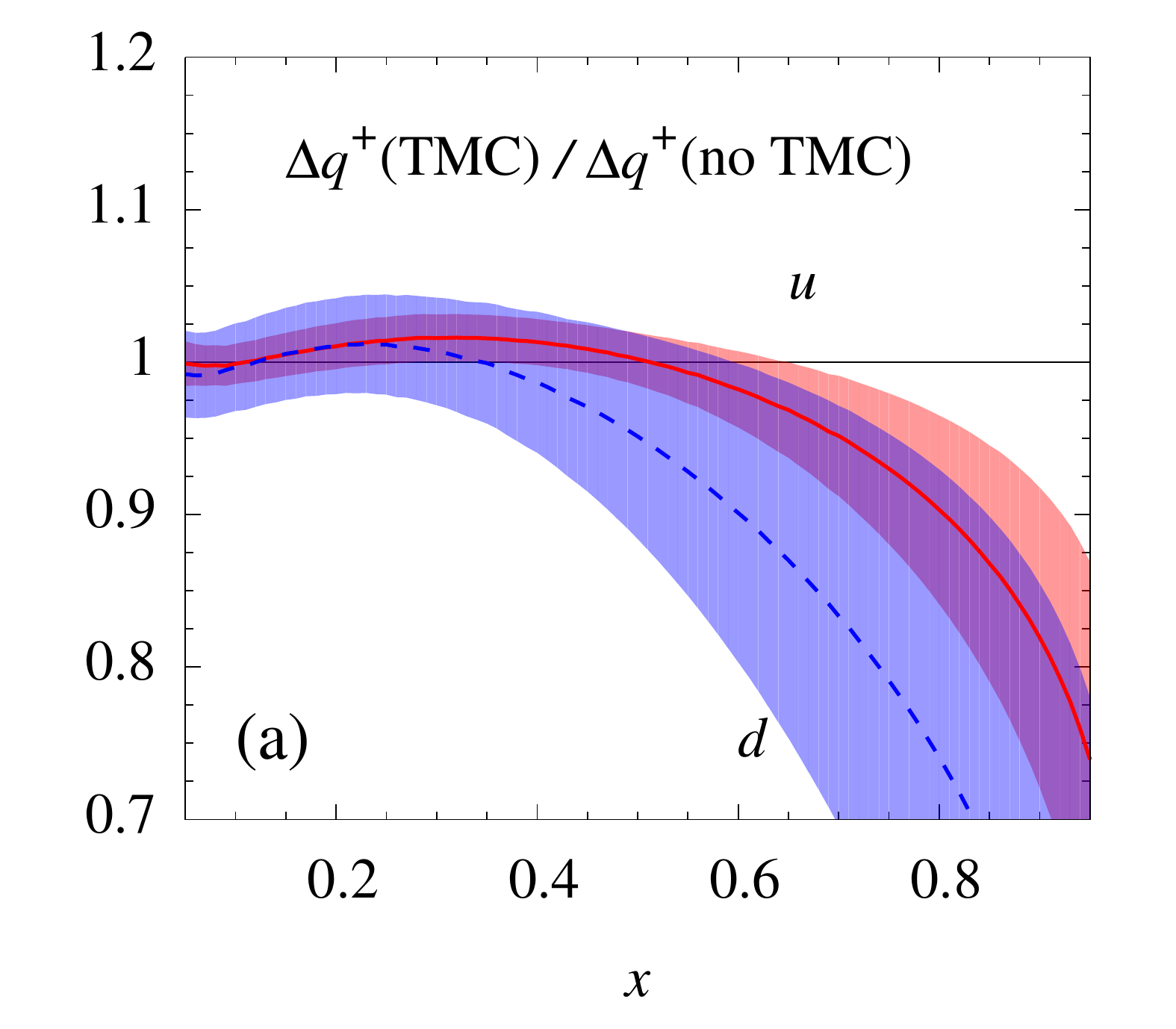}
\includegraphics[width=8cm]{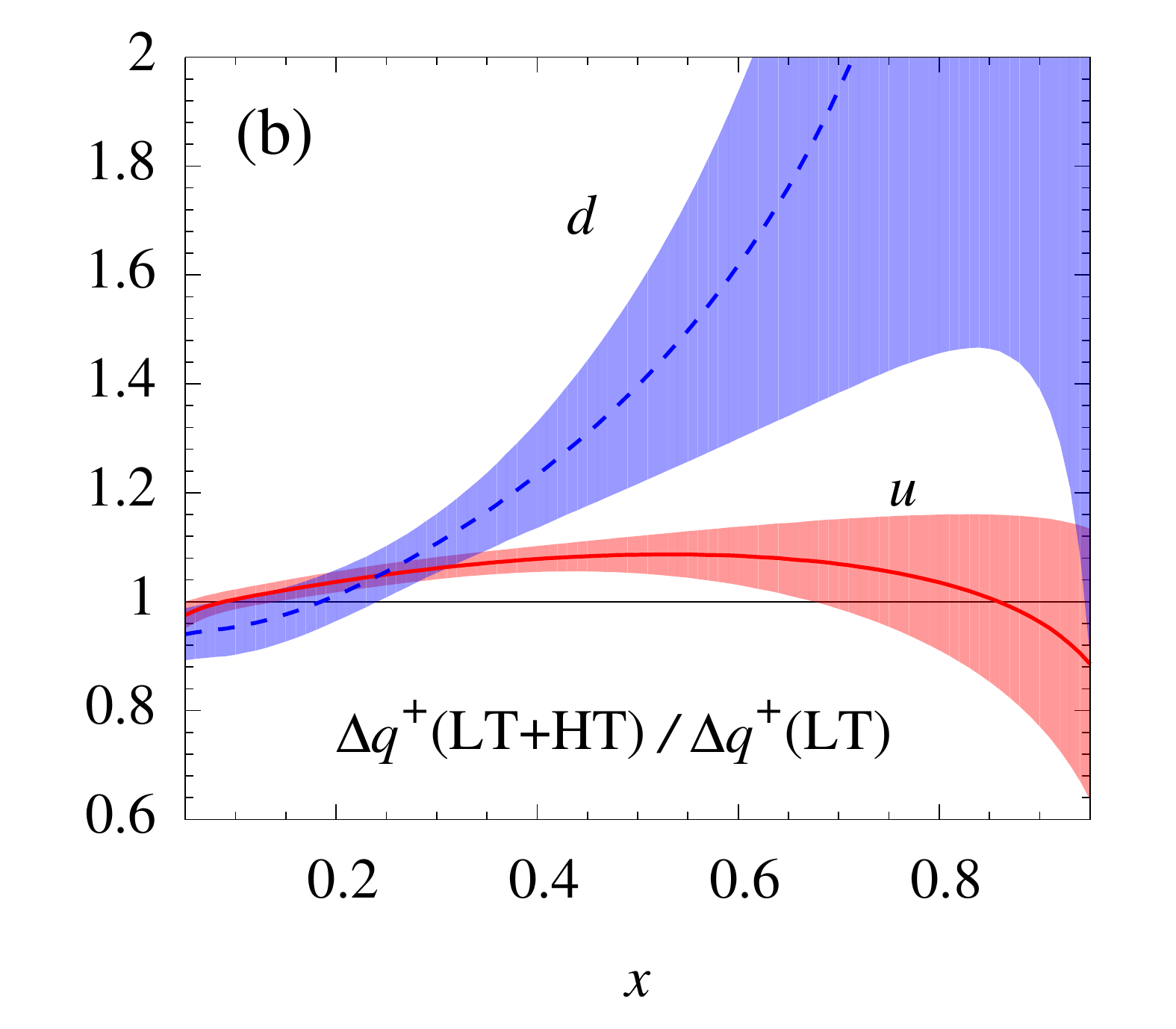}
\caption{
	{\bf (a)}
	Ratio of $\Delta u^+$ and $\Delta d^+$ distributions,
	fitted with target mass corrections, to the reference
	distributions which do not include TMCs.
	All distributions are fitted incorporating nuclear
	smearing corrections.
	{\bf (b)}
	Corresponding ratio of $\Delta u^+$ and $\Delta d^+$
	distributions, computed with higher twist corrections,
	to the leading twist distributions (nuclear smearing
	and target mass corrections are included in both fits).}
\label{fig:finQ2}
\end{figure}

At fixed $Q^2$, the effects of TMCs are most significant at large
values of $x$.  This is illustrated in Fig.~\ref{fig:finQ2}(a)
for the $\Delta u^+$ and $\Delta d^+$ distributions, plotted as a
ratio to the reference leading twist fit from Sec.~\ref{ssec:reffit}
at $Q^2=1$~GeV$^2$, which does not include TMCs.
The effects of the TMCs increase with increasing $x$, with the
distributions including the corrections suppressed relative to
the massless limit fit, particularly for the $\Delta d^+$ PDF.
This is consistent with the enhancement of the $g_1$ structure
function at $x \gtrsim 0.5$ that is generated by the introduction
of TMCs \cite{AM08}; a fit which does not include TMCs will
therefore tend to compensate through larger PDFs in this region.

If one includes subleading $1/Q^2$ corrections through the TMCs,
then for consistency one must also include contributions from
matrix elements of operators with higher twist, $\tau > 2$, which
enter as $Q^2$ power suppressed corrections to cross sections.
These typically involve multi-quark and gluon fields and characterize
elements of the nonperturbative, long-range correlations among partons.
We consider higher twist correction to both the $g_1$ and $g_2$
structure functions,
\begin{subequations}
\label{eq:twist}
\begin{eqnarray}
g_1(x,Q^2)
&=& g_1^{(\tau=2)}(x,Q^2)\
 +\ g_1^{(\tau=3)}(x,Q^2)\ +\ g_1^{(\tau=4)}(x,Q^2),
\label{eq:g1twist}			\\
g_2(x,Q^2)
&=& g_2^{(\tau=2)}(x,Q^2)\ +\ g_2^{(\tau=3)}(x,Q^2),
\label{eq:g2twist}
\end{eqnarray}
\end{subequations}
with the twist-2 contributions given by Eqs.~(\ref{eq:g1LT}) and
(\ref{eq:g2WW}).  The twist-3 contributions to $g_1$ and $g_2$ are
related by the Bl\"umlein-Tkabladze identity \cite{Blumlein:1998nv},
\begin{eqnarray}
g_1^{(\tau=3)}(x,Q^2)
&=& \gamma\
    \Big( g_2^{(\tau=3)}(x,Q^2)
	 - 2 \int_x^1 \frac{dy}{y}\, g_2^{(\tau=3)}(y,Q^2)
    \Big).
\label{eq:g1tw3}
\end{eqnarray}
For the $\tau=3$ part of $g_2$, several parametric forms have been
proposed in the literature.  For the standard JAM analysis we use
a form suggested by the model calculation in Braun {\it et al.}
\cite{BLMP11},
\begin{eqnarray}
g_2^{(\tau=3)}(x,Q^2)
&=& t_0 \left( \ln x + 1-x + \tfrac12 (1-x)^2 \right)   \nonumber\\
& &
 +\ \left( t_1 + t_2(1-x) + t_3(1-x)^2 + t_4(1-x)^3 \right) (1-x)^3,
\label{eq:g2BLMP}
\end{eqnarray}
and fit the coefficients $t_{0 - 4}$ to data.

The parametrization (\ref{eq:g2BLMP}) is relatively flexible,
and reducing the number of parameters does not substantially
change the results.  In practice, therefore, the coefficient
$t_4$ of the highest order $(1-x)$ term is set to zero, $t_4=0$.
Furthermore, we analytically impose that the BC sum rule
(\ref{eq:g2BC}) is satisfied, leaving us with 3 free parameters.
The dependence of the results on the assumptions for the
parametrization of the higher-twist contributions to $g_2$
will be studied in more detail below.

The twist-4 correction to $g_1$,
\begin{eqnarray}
g_1^{(\tau=4)}(x,Q^2) &=& \frac{h(x)}{Q^2},
\label{eq:g1t4}
\end{eqnarray}
is determined phenomenologically, using a spline approximation
for the $x$ dependence of the numerator $h(x)$.
We use knots for the spline at $x=0.1$, 0.3, 0.5 and 0.7,
with the correction constrained to vanish at $x=0$ and 1.
Note that the form (\ref{eq:g2BLMP}) for $g_2^{(\tau=3)}$ and the  
function $h(x)$ in Eq.~\eqref{eq:g1t4} also neglect the possible
$Q^2$ dependence associated with TMCs to these twist-3 and twist-4
functions \cite{Blumlein:1998nv}, which would contribute at
${\cal O}(1/Q^4)$, as well as with the perturbative $\alpha_s$
dependence of the higher-twist matrix elements.
(The QCD evolution of the twist-3 contributions to $g_2$
was considered in Ref.~\cite{Braun01}.)
This is a reasonable approximation given the current precision
of the polarized DIS data.  In practice, the fitted highest twist
corrections ($\tau=4$ term for $g_1$ and $\tau=3$ for $g_2$) also
absorb contributions from yet higher order terms in $1/Q^2$,
as well as from truncations in the perturbative $\alpha_s$ expansion
and other unaccounted for effects (such as threshold resummation
\cite{SGR} and jet mass corrections \cite{JMC, JMC13}).

\begin{figure}[t]
\includegraphics[width=8.4cm]{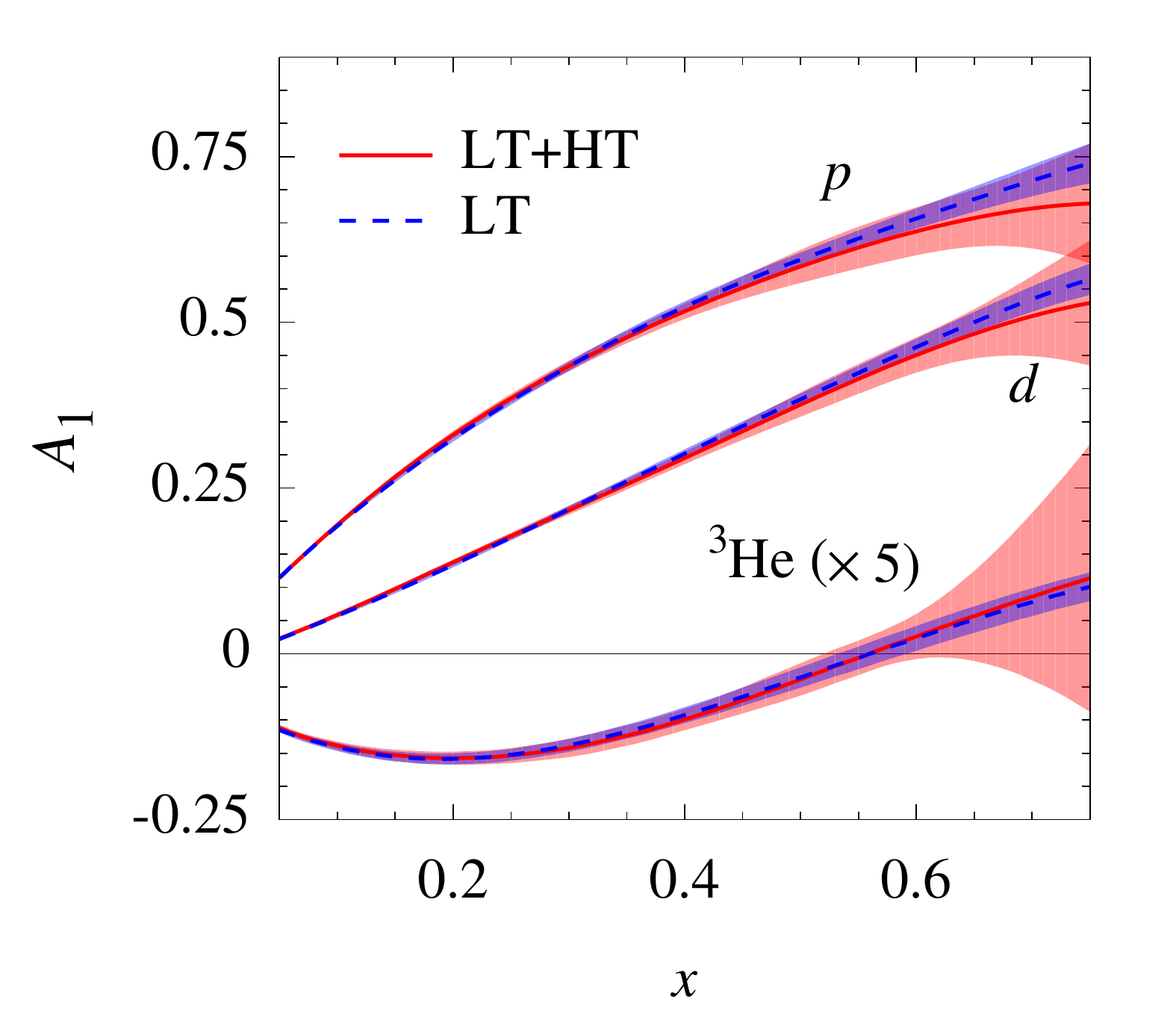}\hspace{-0.6cm}
\includegraphics[width=8.4cm]{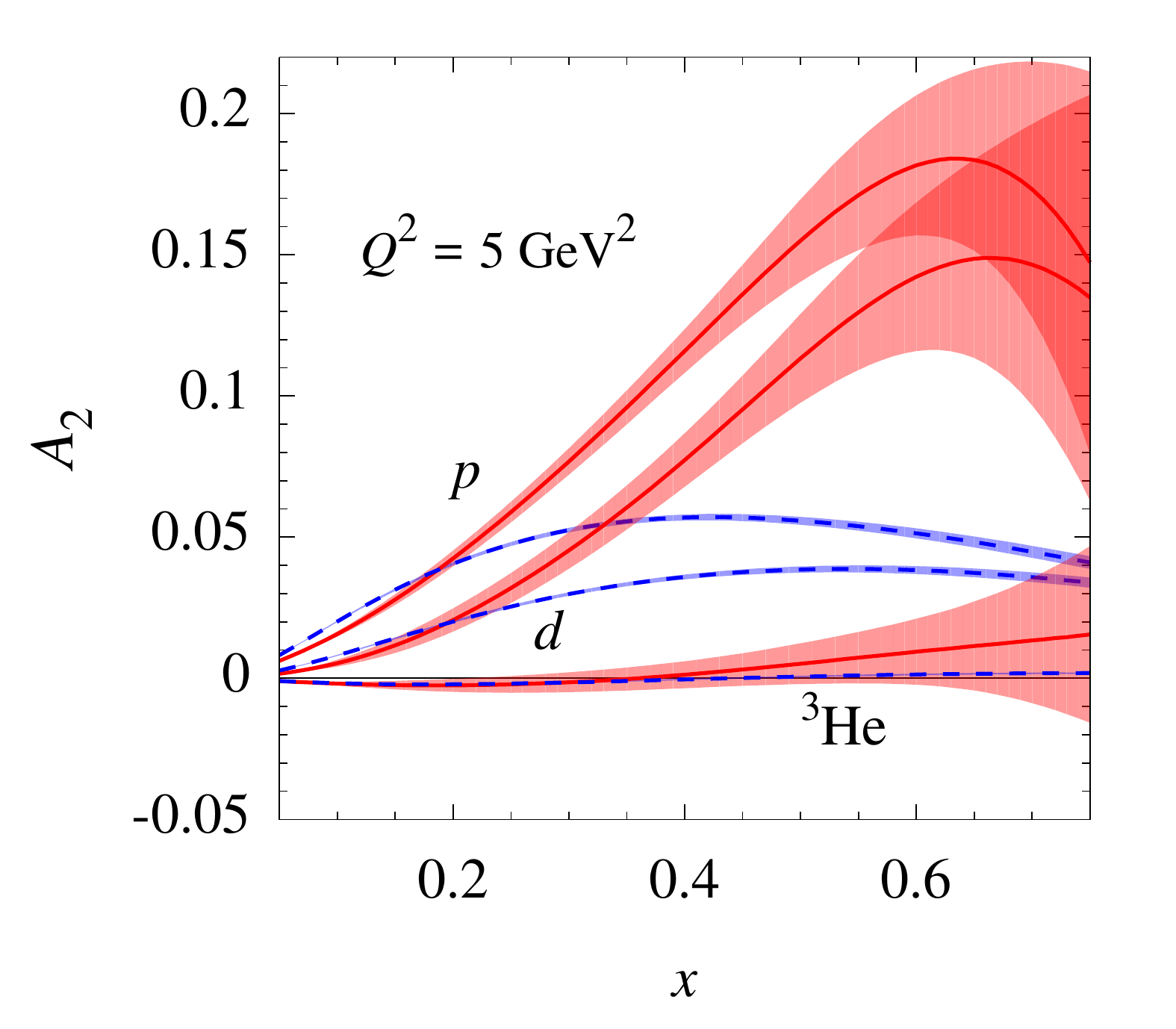}
\caption{Polarization asymmetries $A_1$ ({\bf left}) and $A_2$
	({\bf right}) of the proton, deuteron and $^3$He for the
	JAM fit including leading twist (LT) and higher twist (HT)
	corrections (red solid), and a fit with LT contributions
	only (blue dashed), at $Q^2=5$~GeV$^2$.
	Note that the $^3$He results for $A_1$ are scaled by
	a factor 5 for clarity.}
\label{fig:A1}
\end{figure}

The higher twist corrections have a significant impact on the global
fits, as Fig.~\ref{fig:finQ2}(b) illustrates.  While the $\Delta u^+$
distribution is modified by a modest, $\lesssim 10\%$ correction,
the $\Delta d^+$ PDF is strongly enhanced when higher twists are
taken into account, by more than a factor of 2 for $x \gtrsim 0.6$.
This can be understood by examining the $A_1$ polarization asymmetries
for the proton, deuteron and $^3$He in Fig.~\ref{fig:A1} for the
full JAM fit and for the fit which does not include the higher twist
corrections.
Since the sensitivity to the polarized $d$ quark distribution
is greatest for the neutron $g_1$ data, the $^3$He polarization
asymmetry provides the strongest constraint on $\Delta d^+$.
Note that both the full JAM fit and the fit with leading twist
only give $A_1^{^3{\rm He}}$ asymmetries (along with the proton
and deuteron asymmetries) that are consistent with each other,
and describe the data equally well.  Any nonzero higher twist
component in the full JAM fit would thus have to be offset by an
opposing shift in the leading twist contribution.

\begin{figure}[t]
\includegraphics[width=8.1cm]{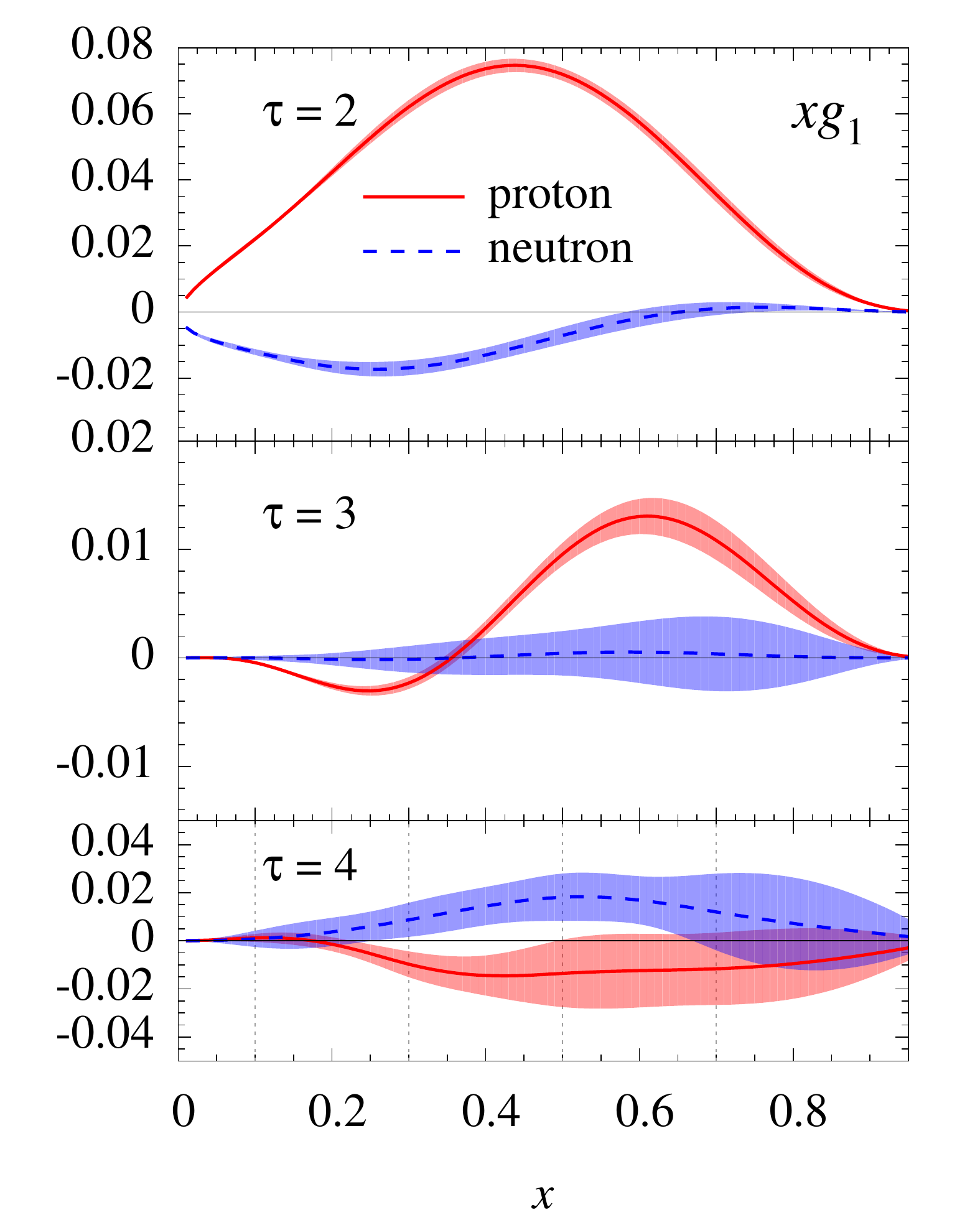}
\includegraphics[width=8.1cm]{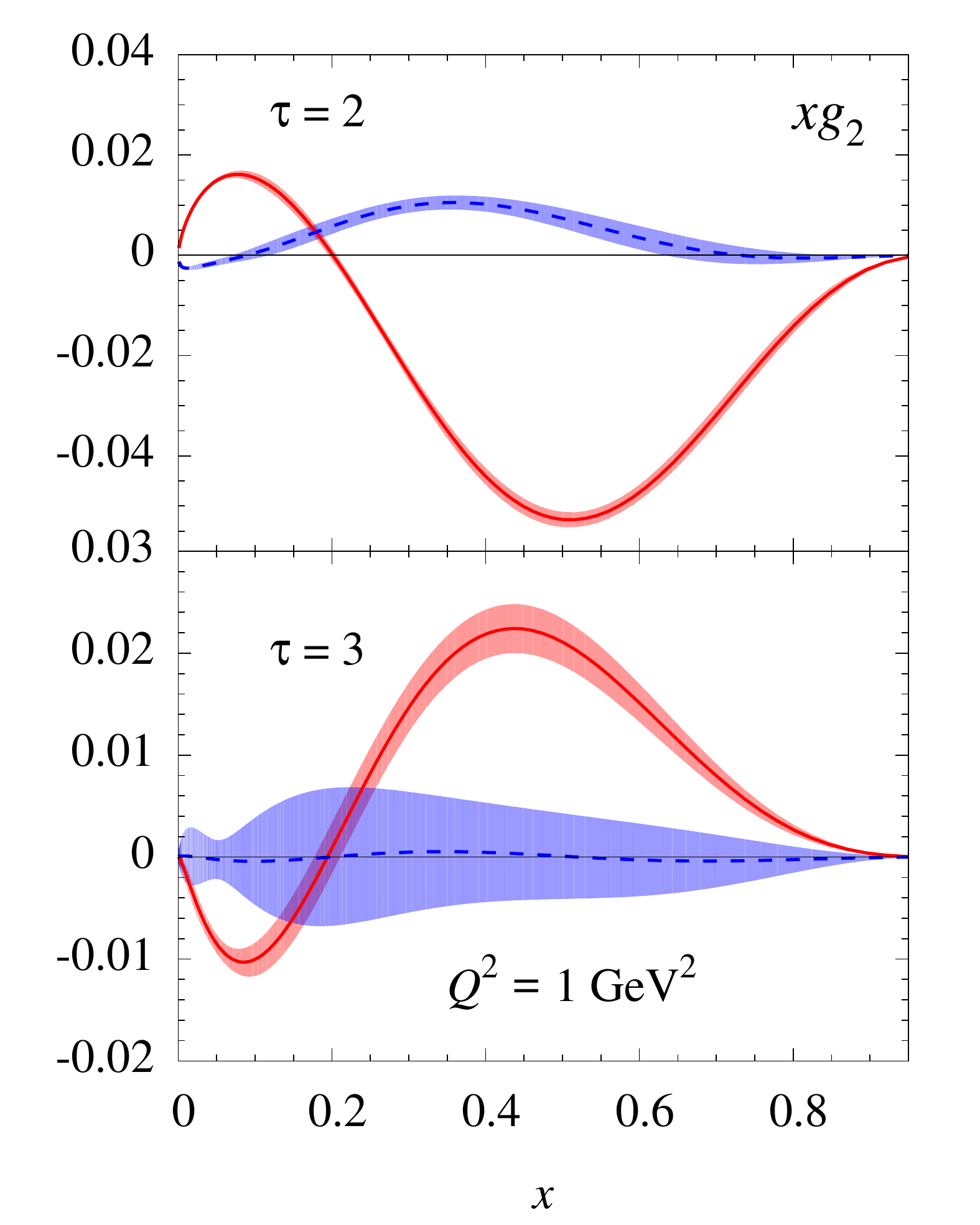}
\caption{Twist decomposition of the proton (red solid) and neutron
	(blue dashed) $xg_1$ and $xg_2$ structure function.
	For the $xg_1$ structure function ({\bf left})
	the twist $\tau=2$ (top panel), $\tau=3$ (middle)
	and $\tau=4$ (bottom) are shown,
	while for the $xg_2$ structure function ({\bf right})
	the $\tau=2$ (top) and $\tau=3$ (bottom) contributions
	are illustrated.
	The dotted vertical lines on the $\tau=4$ contribution
	to $xg_1$ represent the knots used for the spline fit.}
\label{fig:g12HT}
\end{figure}

This is indeed observed in Fig.~\ref{fig:g12HT}, where we present
the individual contributions to the $xg_1$ and $xg_2$ structure
functions from the twist-2 and higher twist terms, for both the
proton and neutron, at $Q^2=1$~GeV$^2$.
For the proton $g_1$ structure function, the twist-3 contribution
is found to be positive at intermediate $x$ values, $x \gtrsim 0.4$,
and negative for the twist-4 correction, though compatible with
zero within the errors.
For the neutron, the twist-3 term is consistent with zero,
but the twist-4 contribution is large and positive at
$0.3 \lesssim x \lesssim 0.7$.
To describe the same experimental $^3$He asymmetry, the leading
twist part of the neutron $g_1$ structure function, and hence the
$\Delta d^+$ distribution, must be more negative.  The ratio of
the (more negative) $\Delta d^+$ PDF from the full JAM fit to
the (less negative) distribution in the leading twist only fit
therefore exceeds unity, as seen in Fig.~\ref{fig:finQ2}(b).

Of course, the $^3$He asymmetry depends also on the proton
contribution, which is dominated by the $\Delta u^+$ PDF.
However, the partial cancellation of the twist-3 and twist-4
contributions to $g_1^p$, and the relatively small overall
magnitude of the higher twist component compared with the much
larger leading twist proton contribution (Fig.~\ref{fig:g12HT}),
means that the impact of the higher twist correction on 
$\Delta u^+$ is minimal (Fig.~\ref{fig:finQ2}(b)).

The leading twist PDFs are of course also indirectly affected by the
$g_2$ structure functions, as is evident from Eq.~(\ref{eq:A1A2}).
For the $g_2$ structure function of the proton, we again find
a significant positive twist-3 correction at $x \gtrsim 0.2$,
which is even larger than the $g_1$ correction, and cancels some
of the (negative) leading twist contribution.  The neutron twist-3
correction to $g_2$, on the other hand, is consistent with zero.
The large positive $\tau=3$ contribution to the proton $g_2$
structure function, which is not $Q^2$-suppressed, is also
responsible for the strong enhancement of the $A_2$ asymmetry
for the proton and deuteron seen in Fig.~\ref{fig:A1}, relative
to the leading twist asymmetry.  For the neutron, the $\tau=3$
correction is consistent with zero, leaving essentially no impact
on the $A_2$ asymmetry for $^3$He.

The shape of the proton twist-3 correction is similar to that found
in the phenomenological analysis in Ref.~\cite{ABMS09}, using the
parametrization 
\begin{eqnarray}
g_2^{(\tau>2)}(x,Q^2)
&=& \alpha_0 (1-x)^{\alpha_1} \Big[ (\alpha_1+2) x - 1 \Big],
\label{eq:g2ABMS}
\end{eqnarray}
which was constructed to vanish at $x=1$ and satisfy the BC sum rule
(\ref{eq:g2BC}).
The parametrization (\ref{eq:g2ABMS}) has a single node, which was
found to be located in a similar position as that in the JAM result
in Fig.~\ref{fig:g12HT}, but with a smaller magnitude.
The result of the calculation of Braun {\it et al.} \cite{BLMP11},
on the other hand, is similar in magnitude to the JAM result,
but of opposite sign at the peak near $x \sim 0.3$.
This may be related to the jet mass corrections, which have sizeable
contributions to $g_2 - g_2^{(\tau=2)}$ starting already at order
$(1/Q^2)^0$ \cite{JMC13}.

\begin{figure}[t]
\includegraphics[width=8.1cm]{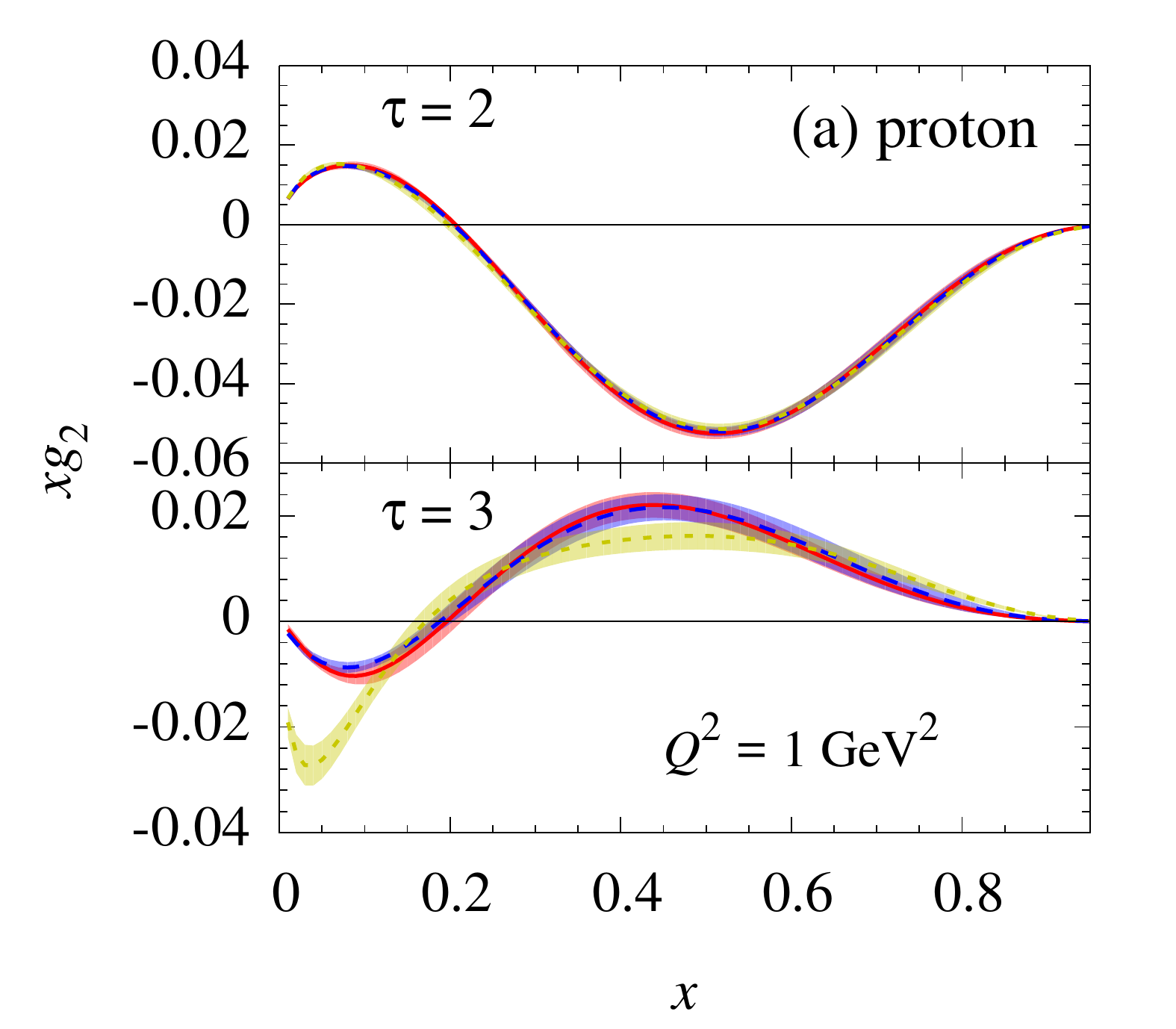}
\includegraphics[width=8.1cm]{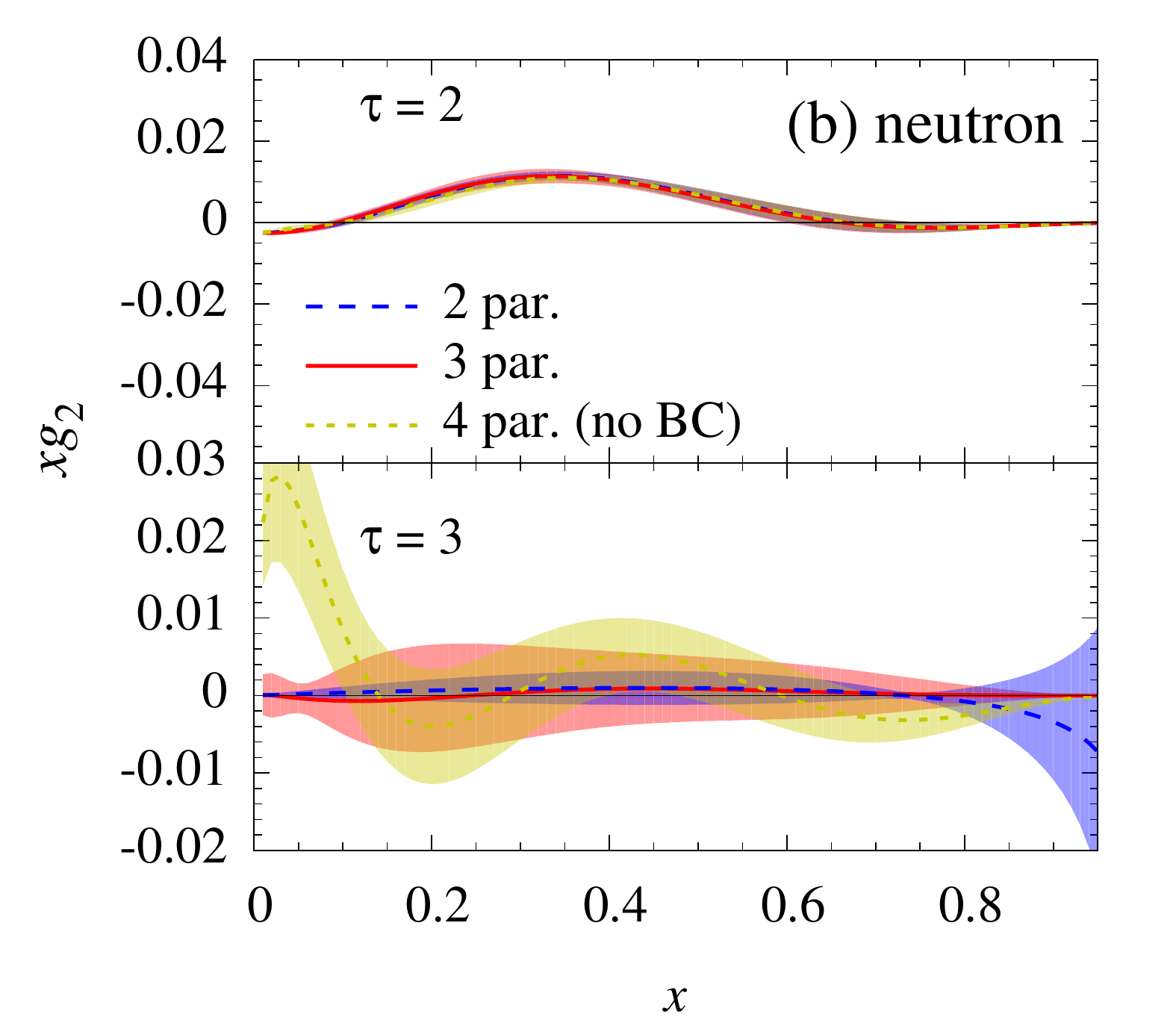}
\caption{Contributions to the $g_2$ structure function of
	{(\bf a)} the proton and 
	{(\bf b)} the neutron
	from the twist $\tau=2$ and $\tau=3$ components.
	The results of the
		3-parameter fit (red solid lines)
	using the functional form in Eq.~(\ref{eq:g2BLMP})
	are compared with the
		2-parameter fit (blue dashed lines)
	of Eq.~(\ref{eq:g2ABMS}), and with a
		4-parameter fit (yellow dotted lines)
	employing Eq.~(\ref{eq:g2BLMP}) but without imposing
	the BC sum rule constraint (\ref{eq:g2BC}).}
\label{fig:g2HT}
\end{figure}

To study the model dependence of the extracted higher twist correction
to $g_2$, we perform an additional fit of the data using the functional
form in Eq.~(\ref{eq:g2ABMS}), but with the parameters $\alpha_0$ and
$\alpha_1$ refitted to the extended data used in the JAM analysis.
The result of the refit using (\ref{eq:g2ABMS}) is remarkably similar
to the JAM higher twist correction, for both the proton and neutron,
as illustrated in Fig.~\ref{fig:g2HT}, which suggests that the different
magnitude found in Ref.~\cite{ABMS09} was driven by the input data.
As a further check, we fit the data using the full JAM fit form
(\ref{eq:g2BLMP}), but do not impose the BC sum rule constraint
(\ref{eq:g2BC}).
Once again the result is very similar at all values of $x \gtrsim 0.1$,
with the unconstrained (negative) proton correction slightly larger
in magnitude at smaller $x$.  The unconstrained neutron correction is
positive at small $x$, but is compatible with zero at larger $x$ values.
The values of the lowest $g_2$ moments from the unconstrained fit are
found to be $g_2^{(1), {\rm fit}} = -0.071 \pm 0.012$ for the proton
and $0.092 \pm 0.036$ for the neutron.
Overall, this suggests that the extraction of the higher twist
contribution to $g_2$ is not strongly dependent on the assumed
parametrization, and points to a relatively small violation of the
BC sum rule.

An additional window on higher twist dynamics is afforded by the
$d_2$ matrix element of the nucleon, which measures a specific
combination of the $n=3$ moments of the $g_1$ and $g_2$ structure
functions, $d_2(Q^2) = 2 g_1^{(3)}(Q^2) + 3 g_2^{(3)}(Q^2)$.
At leading twist this combination of moments vanishes, so that
measurement of this matrix elements reveals, to leading order in
$1/Q^2$, the twist-3 contribution to $g_2$.  Our global fits give
for the proton $d_2$ matrix element
  $d_2^p = 0.011 \pm 0.002$
at a scale of $Q^2 = 5$~GeV$^2$, and for the neutron
  $d_2^n = 0.002 \pm 0.003$.
The SLAC E155x experiment measured the $d_2$ moments
to be \cite{SLAC-E155x}
  $d_2^{p (\rm exp)} = 0.0032 \pm 0.0017$ and
  $d_2^{n (\rm exp)} = 0.0079 \pm 0.0048$
at an average $Q^2 = 5$~GeV$^2$, which are in general
agreement with the global JAM values within the uncertainties
(note, however, that the $\alpha_s$ dependence of the twist-3
part of $d_2$ is currently not taken into account in the JAM
analysis).
More recently, the RSS experiment \cite{RSS07} in Jefferson Lab
Hall~C obtained the contribution
  $d_2^{p (\rm exp)} = 0.0057 \pm 0.0011$
from the resonance region $0.29 < x < 0.84$ at
$\langle Q^2 \rangle = 1.3$~GeV$^2$,
where the experimental errors have been added in quadrature.
Including the extrapolation into the unmeasured region, the
inelastic contributions to the proton and neutron moments
were found to be \cite{RSS10}
  $d_2^{p (\rm exp)} = 0.0104 \pm  0.0014$
and
  $d_2^{n (\rm exp)} = -0.0075 \pm 0.0021$
at $\langle Q^2 \rangle = 1.28$~GeV$^2$.
The contribution from the resonance region to the neutron $d_2$
was also recently measured in the E01-012 experiment in Hall~A
to be \cite{E01-012g2}
  $d_2^{n (\rm exp)} = 0.0002 \pm 0.0010$
at $\langle Q^2 \rangle = 2.4$~GeV$^2$.
The E06-014 experiment \cite{E06-014} in Jefferson Lab Hall~A is
currently analyzing the data on the $d_2$ matrix element of $^3$He,
from which the neutron value will be extracted at
$\langle Q^2 \rangle \approx 4$~GeV$^2$.

%%%%%%%%%%%%%%%%%%%%%%%%%%%%%%%%%%%%%%%%%%%%%%%%%%%%%%%%%%%%%%%%%%%%%%%%%
\section{JAM fit results}
\label{sec:JAMfits}

\begin{figure}[t]
\includegraphics[width=8cm]{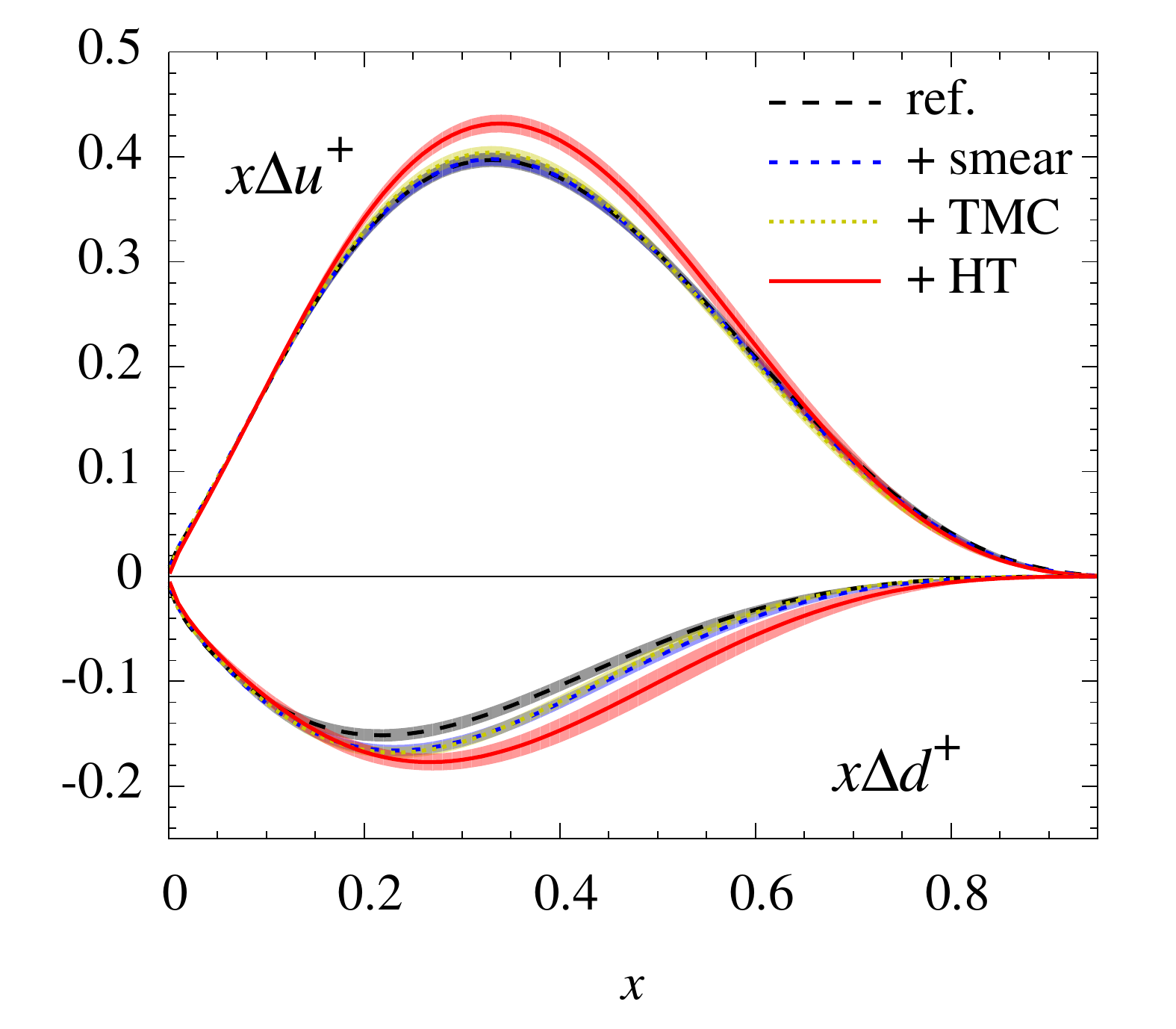}
\includegraphics[width=8cm]{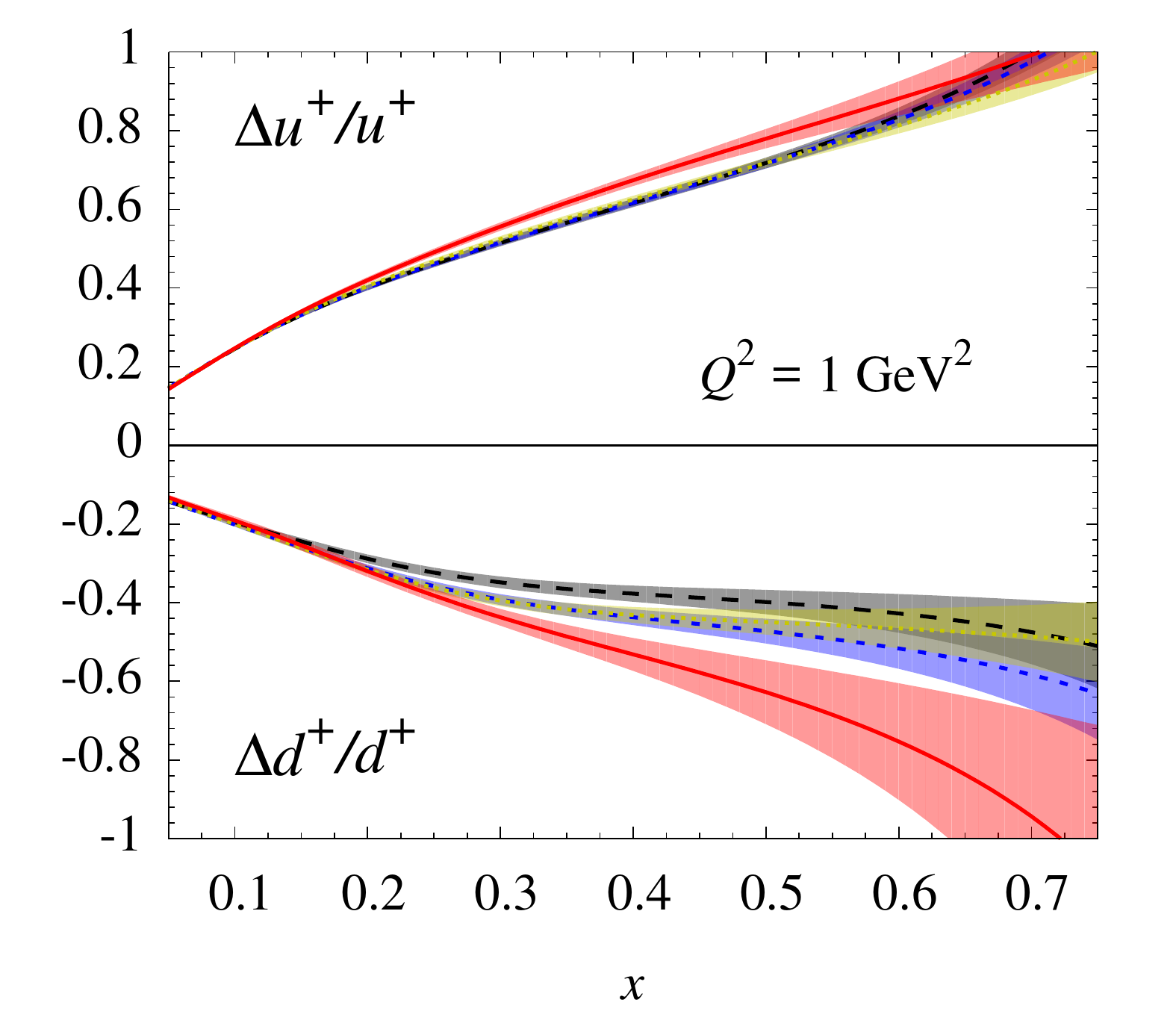}
\caption{{\bf (Left)}
	Spin-dependent $x\Delta u^+$ and $x\Delta d^+$ distributions
	showing the cumulative effects on the reference PDFs
	(black dashed line) of adding nuclear smearing
	(blue short-dashed), target mass (green dotted),
	and higher twist (red solid) corrections.
	{\bf (Right)}
	Corresponding ratios of polarized to unpolarized
	$\Delta u^+/u^+$ and $\Delta d^+/d^+$ distributions.}
\label{fig:cumulative}
\end{figure}

Combining the effects of the nuclear and finite-$Q^2$ corrections
discussed in the previous sections, the results of the full JAM
fits to the data sets in Table~\ref{tab:data} are summarized in
Fig.~\ref{fig:cumulative}.  Beginning with the reference
parametrizations defined in Sec.~\ref{ssec:reffit}, the cumulative
effects of the nuclear smearing, target mass and higher twist
corrections on the $\Delta u^+$ and $\Delta d^+$ distributions
are demonstrated explicitly.
(Since the reference fit here is the same as that in Fig.~\ref{fig:ref},
the differences between the JAM fits for $\Delta u^+$ and $\Delta d^+$
and the results from previous PDF analyses can also be compared.)

The impact of these corrections is negligible at small values of $x$,
$x \lesssim 0.2$, but grows increasingly important at higher $x$.
Compared with the reference distributions, both the JAM $\Delta u^+$
and $\Delta d^+$ PDFs are larger in magnitude, by $\sim 10\%-20\%$
for the $u$ quark at $0.2 \lesssim x \lesssim 0.6$, and by more
than $50\%-100\%$ for the $d$ quark at $x \gtrsim 0.4$.
The same effects are more clearly illustrated in the form of
ratios of polarized to unpolarized PDFs $\Delta u^+/u^+$ and
$\Delta d^+/d^+$, shown in Fig.~\ref{fig:cumulative}(b).
Such a comparison is meaningful since the unpolarized PDFs are
fitted within the same analysis and applied consistently in
the extraction of the polarized PDFs.

At intermediate $x$ values, both the $u$ and $d$ quark ratios
are broadly consistent with the SU(6) quark model predictions
of $\Delta u/u = 2/3$ and $\Delta d/d = -1/3$ \cite{MT96}.
At larger $x$, the $\Delta u^+/u^+$ ratio continues to rise,
as expected from helicity conservation models and perturbative
QCD calculations \cite{Farrar:1975yb}, which predict that for
all quark flavors the ratio $\Delta u^+/u^+ \to 1$ as $x \to 1$.
The $\Delta d^+/d^+$ ratio, however, remains negative with no
indication of the upturn predicted by the helicity arguments.
Clearly, the nuclear and finite-$Q^2$ effects have significantly
impact on the asymptotic $x \to 1$ behavior, and additional data
are vital to provide constraints at high $x$.

\begin{table}[pt]
\begin{center}
\caption{Parameter values for the leading twist JAM
	distributions at the input scale $Q^2 = 1$~GeV$^2$.
	The fitted values are those including errors, while
	the parameters shown without errors are determined
	through the relations in Sec.~\ref{ssec:param}.
	Note that the leading twist contributions need
	to be complemented by the higher twist terms
	in Tables~\ref{tab:HTg1} and \ref{tab:HTg2}.}
\begin{tabular}{c|c|c|c|c|c}			\hline\hline
flavor $f$\
  & $N_f$
  & $a_f$
  & $b_f$
  & $c_f$
  & $d_f$					\\ \hline
$\Delta u^+$
  &\ \ \ $1.1167 \pm 0.17$\ \
  &\ $0.8244 \pm 0.04$\ \
  &\ $3.3244 \pm 0.12$\ \
  &\ $-1.8722 \pm 0.30$\ \
  &\ $11.2858 \pm 0.8466$\ \			\\
$\Delta d^+$
  & $-0.8374 \pm 0.17$
  &\ $0.7193 \pm 0.05$\ \
  &\ $3.9932 \pm 0.40$\ \
  &\ $-1.9324 \pm 0.40$\ \
  &\ \  $7.0703 \pm 0.7407$			\\
$\Delta g$
  & $-0.8120 \pm 0.90$
  & 1.7718
  & 5.6588
  & 0
  & $-33.8287 \pm 21$				\\
$\Delta\bar{u}$
  &\ \ \ 0.5583 & 0.8244 & 7.6588 & 0 & 0	\\
$\Delta\bar{d}$
  & $-0.4186$ & 0.7193 & 7.6588 & 0 & 0 	\\
$\Delta\bar{s}$
  & $-0.2073 \pm 0.04$
  & 0.7193 & 7.6588 & 0 & 0			\\ \hline
\end{tabular}
\label{tab:parameters}
\end{center}
\begin{center}
\caption{Parameter values for the coefficient $h(x)$ of the twist
	$\tau=4$ contribution to the $g_1$ structure function in
	Eq.~(\ref{eq:g1t4}) at $Q^2 = 1$~GeV$^2$.  The function
	values are given at the knots $x=0.1$, 0.3, 0.5 and 0.7.}
\begin{tabular}{l|c|c|c|c}				\hline\hline
\ \ $g_1^{(\tau=4)}$
	&      $h(0.1)$	     &   $h(0.3)$
	&      $h(0.5)$      &   $h(0.7)$		\\ \hline
proton
	&\ $0.0118 \pm 0.017$\   &$-0.0325 \pm 0.018$\
	&\ $-0.0271\pm 0.028$\   & $-0.0167 \pm 0.022$\	\\
neutron\ \ 
	&\ $0.0079 \pm 0.034$\   &\ \ $0.0290 \pm 0.024$\
	&\ \ $0.0362 \pm 0.020$\ &\ \ $ 0.0171 \pm 0.023$\ \\ \hline
\end{tabular}
\label{tab:HTg1}
\end{center}
\begin{center}
\caption{Parameter values for the coefficients $t_{0-4}$ of the
	twist $\tau=3$ correction to the $g_2$ structure function
	in Eq.~(\ref{eq:g2BLMP}) at $Q^2 = 1$~GeV$^2$.
	The fitted values are shown with error bars.}
\begin{tabular}{l|c|c|c|c|c}		\hline\hline
\ \ $g_2^{(\tau=3)}$
  & $t_0$
  & $t_1$
  & $t_2$
  & $t_3$
  &\ \ \ \ $t_4$\ \ \ \ 		\\ \hline
proton
  &\ $-0.0936$\ \
  &\ \ $0.2837 \pm 0.18$\
  &\ $0.7542 \pm 0.50$\ \
  &\ $-1.5177 \pm 0.4$\	\		
  &\ 0\ 				\\
neutron\ \
  &\ $-0.0193$\ \
  &\ $-0.0136 \pm 0.42$\ \
  &\ $0.1062 \pm 1.25$\ \
  &\ $-0.1456 \pm 1.1$\	\		
  &\ 0\ 				\\ \hline
\end{tabular}
\label{tab:HTg2}
\end{center}
\end{table}

Although not utilized in the main JAM analysis, additional data
at low $Q^2$ and low $W^2$ do exist from several Jefferson Lab
experiments \cite{E97-103, EG1b, E01-012g2, E01-012g1}, which could
affect the large-$x$ behavior of the PDFs.  In the next section we
explore in greater detail the effect on the global fits of these
data in order to assess the possible implications of including
low-$W$ and low-$Q^2$ data in future global analyses.

For completeness, the parameter values for the full JAM fit to
the leading twist parton distributions are presented in
Table~\ref{tab:parameters} at the input scale $Q^2 = 1$~GeV$^2$.
The parameters for the twist-4 correction to the $g_1$ structure
function in Eq.~(\ref{eq:g1t4}) and the twist-3 correction to
the $g_1$ structure function in Eq.~(\ref{eq:g1t4}) are listed
in Tables~\ref{tab:HTg1} and \ref{tab:HTg2}, respectively.
In total we find that the inclusive DIS data are able to
constrain 13 leading twist parameters, and 7 higher twist
parameters each for the proton and neutron.
(The fitted parameters in Tables~\ref{tab:parameters}
-- \ref{tab:HTg2} are given with error bars.)
The remaining parameters are constrained using the relations in
Secs.~\ref{ssec:param} and \ref{sec:finiteQ}.
Note that the values in the tables are given to more significant
figures than would be appropriate given the quoted error,
in order to accurately reproduce our fitted distributions
in numerical calculations.

We stress also that the leading twist parameters should {\it not}
be used to reconstruct the structure functions or polarization
asymmetries in isolation of the higher twist contributions.
Due to the importance of the higher twist corrections in these
fits, attempts to reproduce the data, or making predictions for
future measurements, with the leading twist distributions alone
would lead to discrepancies with experiment.

%%%%%%%%%%%%%%%%%%%%%%%%%%%%%%%%%%%%%%%%%%%%%%%%%%%%%%%%%%%%%%%%%%%%%%%%%
\section{Data selection and cut sensitivity}
\label{sec:cuts}

One of the central and novel features of the present JAM global
analysis is the inclusion of data over an extended range of
kinematics, $Q^2 \geq 1$~GeV$^2$ and $W^2 \geq 3.5$~GeV$^2$,
together with the theoretical corrections needed to reliably
account for the nuclear and finite-$Q^2$ effects that form an
integral part of such an analysis.
The cuts themselves are similar to those adopted in some previous
analyses; however, the various corrections have not always been
applied uniformly.
For example, the DSSV analysis \cite{DSSV09} applies a cut
of $Q^2 > 1$~GeV$^2$, but extracts the $A_1$ asymmetry in
Eq.~(\ref{eq:A1A2}) using the approximation
	$A_1 \approx A_1^{(0)} / (1+\gamma^2)$,
where $A_1^{(0)}$ is the asymmetry in the massless
($M^2/Q^2 \to 0$) limit, which neglects the $g_2$ contribution
as well as higher orders in the $1/Q^2$ expansion of the TMCs.
The BB fit \cite{BB10}, on the other hand, uses a similar cut in
$Q^2$ and implements TMCs more systematically, but does not apply
nuclear smearing corrections despite using a weaker $W$ cut,
$W^2 > 3.24$~GeV$^2$.  In a more recent analysis \cite{BB12},
BB also consider higher twist corrections to $g_2$, pointing
out the need for $Q^2$ evolution of the twist-3 term
\cite{BLMP11, Braun01}.
As mentioned in Sec.~\ref{sec:finiteQ} above, the possible scale
dependence of the twist-3 (and twist-4) contributions is not
considered in the current JAM fit, but will be included in a
subsequent analysis \cite{JAM-III}.

In this section we investigate the stability of the JAM fits
with respect to variations in the $Q^2$ and $W^2$ cuts from
their nominal values, incorporating {\it a priori} the full TMC,
higher twist, and nuclear smearing corrections.
In particular, we consider the effects of varying the $W^2$
cut between 3 and 4~GeV$^2$, as well as increasing the $Q^2$
cut to 2~GeV$^2$.  In addition, we perform a leading twist
fit utilizing the more stringent cut of $W^2 > 6.25$~GeV$^2$,
as advocated in Ref.~\cite{NNPDF13} to avoid finite-$Q^2$
corrections.

\begin{figure}[t]
\includegraphics[width=8cm]{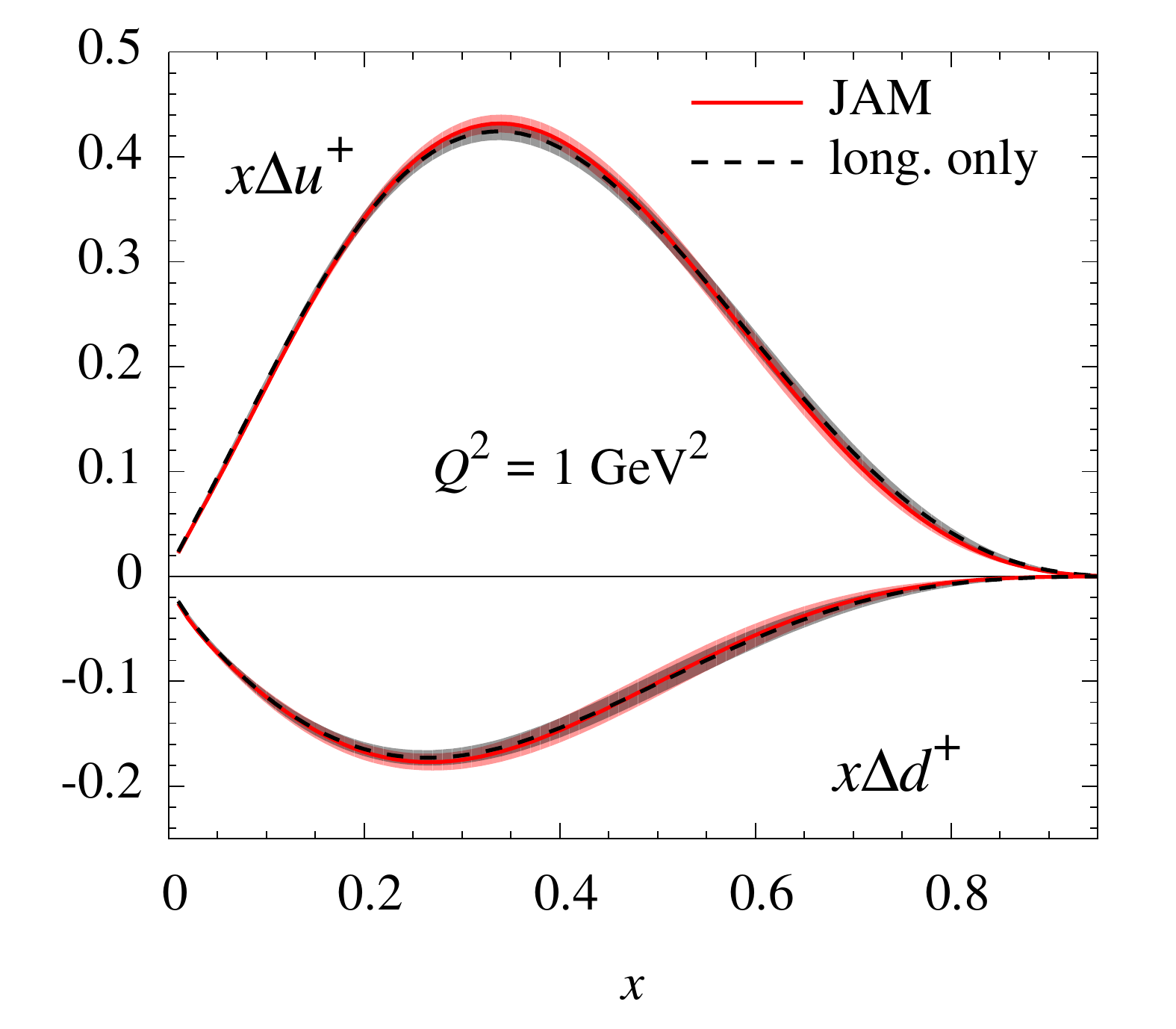}
\includegraphics[width=8cm]{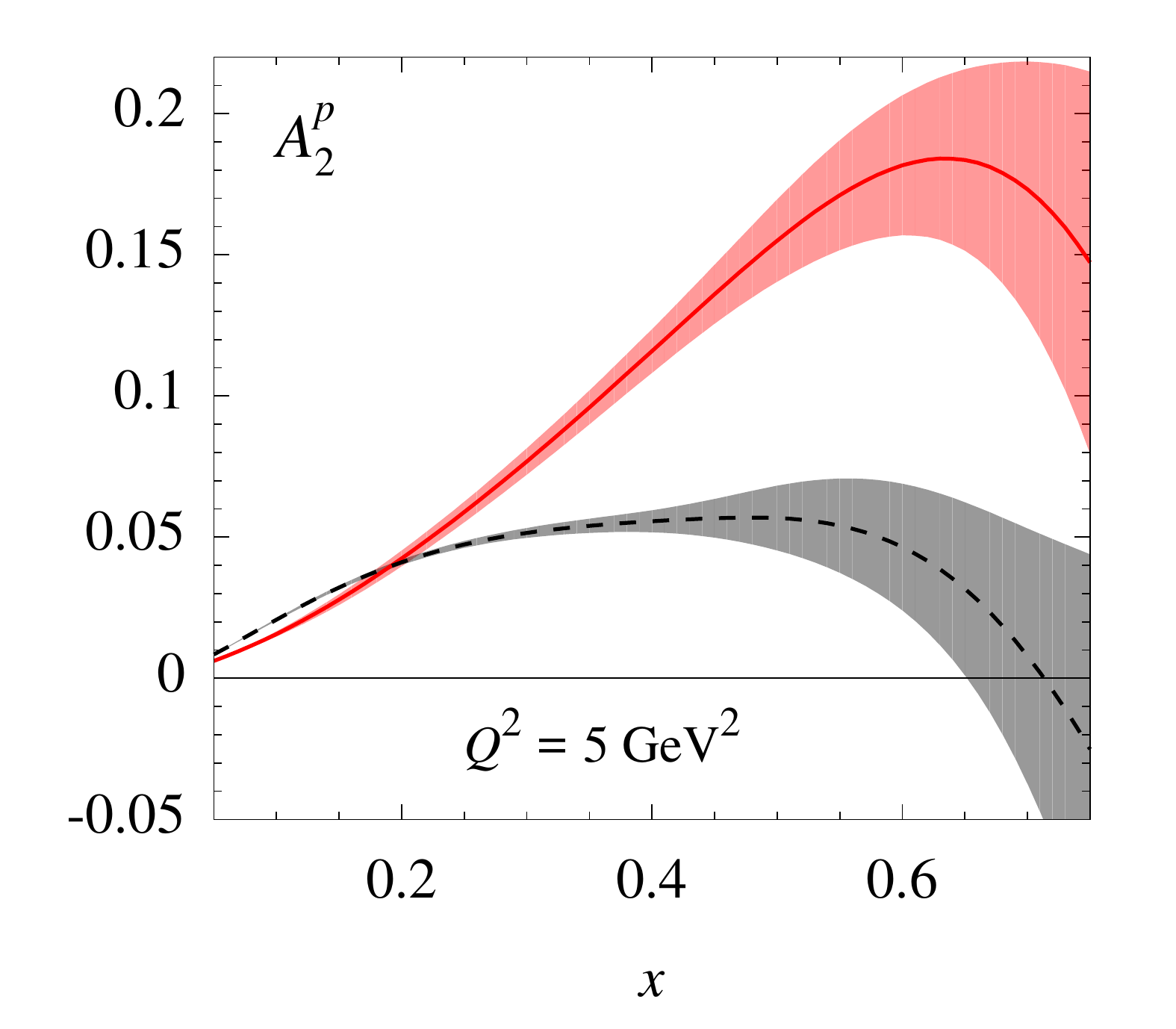}
\caption{Comparison of the full JAM fit (red solid lines) with a fit
	using only longitudinal asymmetries $A_\parallel$ (or $A_1$)
	(black dashed), for the
	{\bf (left)}
	$x\Delta u^+$ and $x\Delta d^+$ distributions
	at $Q^2=1$~GeV$^2$, and
        {\bf (right)}
	proton $A_2^p$ asymmetry at $Q^2=5$~GeV$^2$.}
\label{fig:Lonly}
\end{figure}

Before examining the cut dependence, however, we first consider the
effects of including in the global fit data on both longitudinal
and transverse asymmetries (or on $A_1$ and $A_2$ in cases where
$A_\parallel$ and $A_\perp$ are not available).  The strategy that
is commonly employed in many other analyses of spin-dependent
PDFs is to fit $A_1$ (or $g_1$) data extracted from the measured
asymmetries while fixing the $g_2$ structure function.
For example, the standard assumption is to approximate $g_2$ by
the twist-2, Wandzura-Wilczek contribution [Eq.~(\ref{eq:g2WW})],
which also implies a vanishing twist-3 contribution to $g_1$
[Eq.~(\ref{eq:g1tw3})].

The differences between the full JAM analysis, which simultaneously
fits $A_\parallel$ and $A_\perp$ data, and that based on longitudinal
asymmetries only are illustrated in Fig.~\ref{fig:Lonly}.
Since the differences between $A_1$ and $A_\parallel$ lie mainly
(although not exclusively) in the higher twist content of the spin
structure functions, the impact on the $\Delta u^+$ and $\Delta d^+$
distributions is minimal, with the phenomenological twist-4 correction
to $g_1$ absorbing most of the differences.
On the other hand, the effect on the higher twist contributions
to the spin structure functions, particularly $g_2$, is significant,
as may be anticipated given that longitudinal asymmetries generally
receive very small contributions from $g_2$.
The typical impact of the transverse asymmetry data on the global
fits is illustrated in Fig.~\ref{fig:Lonly} for the proton $A_2$
asymmetry, which shows a signicant (factor $\approx 2-3$) enhancement
in the full fit at $x \gtrsim 0.4$.
One may conclude therefore that while the simultaneous fit gives
similar results for the twist-2 PDFs as one based on longitudinal
data only, transverse asymmetry data are essential if one is to
determine in addition the higher twist content of the structure
functions.

\begin{figure}[t]
\includegraphics[width=8cm]{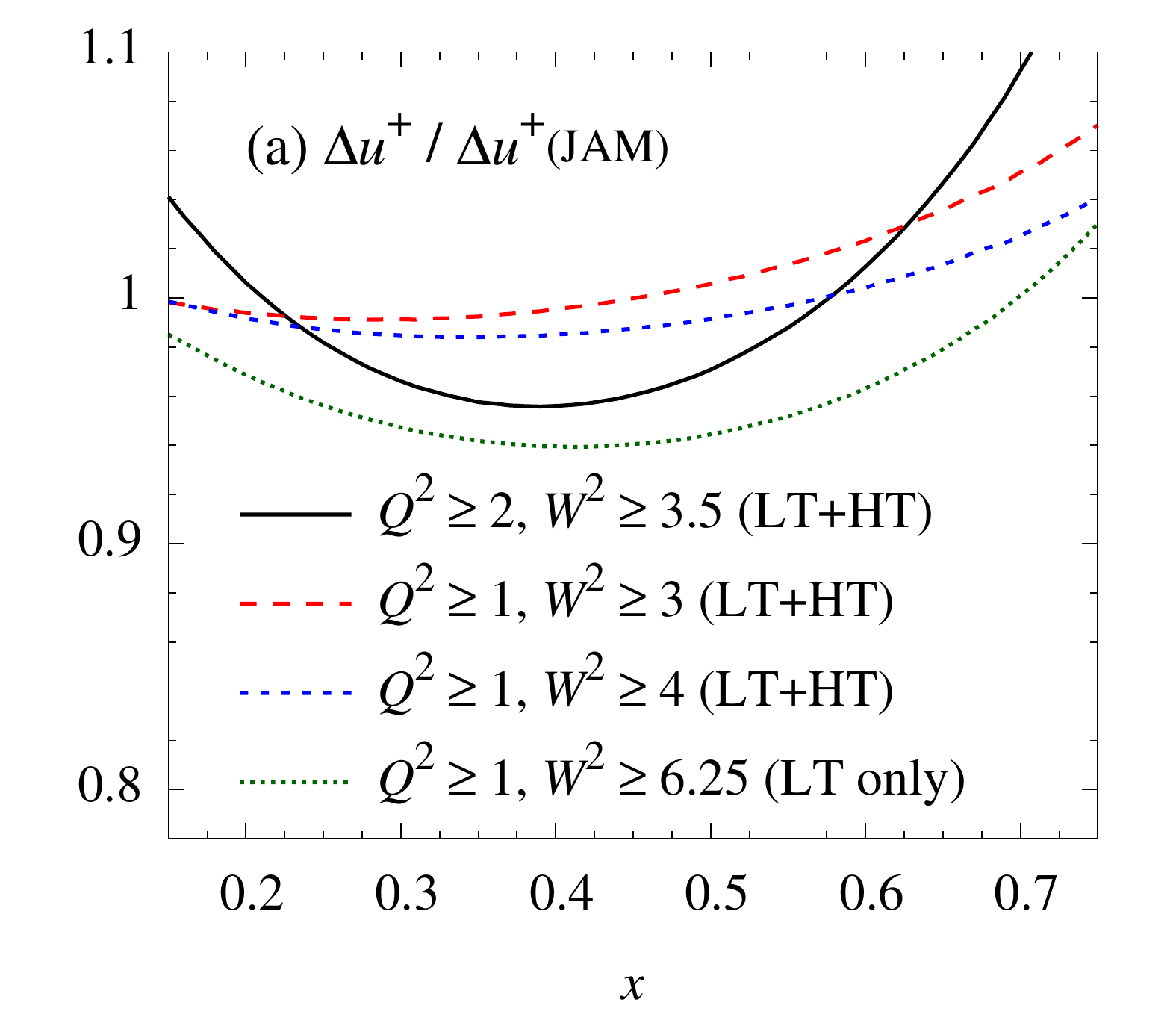}
\includegraphics[width=8cm]{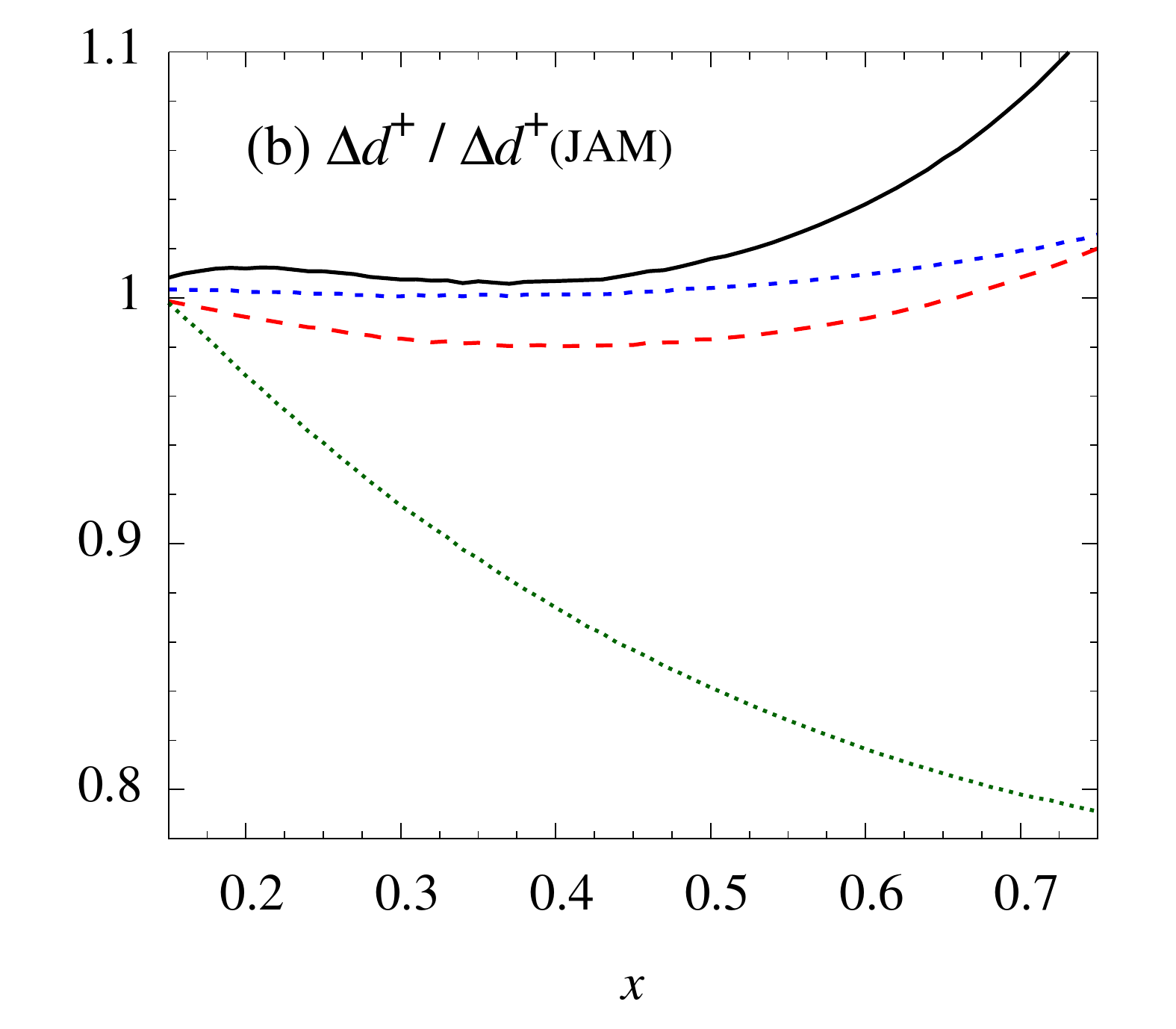}
\caption{Ratio of PDFs with various $Q^2$ and $W^2$ cuts,
	to the JAM PDFs with the nominal
	$Q^2 \geq 1$~GeV$^2$ and $W^2 \geq 3.5$~GeV$^2$ cuts, for
	{\bf (a)} the $\Delta u^+$ and
	{\bf (b)} the $\Delta d^+$ distribution.
	The variations include the cuts
	$Q^2 \geq 2$~GeV$^2$, $W^2 \geq 3.5$~GeV$^2$ (black solid),
	$Q^2 \geq 1$~GeV$^2$, $W^2 \geq 3$~GeV$^2$ (red dashed), and
	$Q^2 \geq 1$~GeV$^2$, $W^2 \geq 4$~GeV$^2$ (blue short-dashed)
	utilizing LT and HT corrections, and
	$Q^2 \geq 1$~GeV$^2$, $W^2 \geq 6.25$~GeV$^2$ (green dotted)
	with LT only.}
\label{fig:cuts}
\end{figure}

The effect of varying the $Q^2$ and $W^2$ cuts on the $\Delta u^+$
and $\Delta d^+$ PDFs is illustrated in Fig.~\ref{fig:cuts} at a
scale of $Q^2=1$~GeV$^2$.  Compared with the nominal JAM cuts of
$Q^2 \geq 1$~GeV$^2$ and $W^2 \geq 3.5$~GeV$^2$, the variation
of the $W^2$ cut between 3~GeV$^2$ and 4~GeV$^2$ has a small,
$\lesssim 2\%$ effect on both the $\Delta u^+$ and $\Delta d^+$
distributions, compared with the PDF uncertainties, for most of
the $x$ range shown.  This suggests that extending the kinematic
reach of inclusive DIS data to (marginally) inside the traditional
nucleon resonance region ($W \lesssim 2$~GeV) can still yield stable,
leading twist distributions, provided finite-$Q^2$ corrections are
taken into account.  This finding has previously also been observed in
global analyses of unpolarized PDFs \cite{CJ12, CJ10, CJ11, ABKM09}.
Qualitatively similar effects are observed when the $Q^2$ cut
is varied, from 1 to 2~GeV$^2$, with $\Delta u^+$ changing by
$\lesssim 4\%$ for $x < 0.6$, and $\Delta d^+$ by $\sim 1\%-2\%$
for $x < 0.5$.  At higher $x$, both the $\Delta u^+$ and
$\Delta d^+$ PDFs are enhanced by $\sim 10\%$ for $x \sim 0.7$,
which is outside of the region directly constrained by data.
Note, however, that for $W^2 \geq 3.5$~GeV$^2$, the restriction to
$Q^2 \geq 1$~GeV$^2$ means that only the region $x \lesssim 0.28$ is
directly constrained by data at this scale, and $x \lesssim 0.43$
for $Q^2 \geq 2$~GeV$^2$.  In this region the variation in both the
$\Delta u^+$ and $\Delta d^+$ PDFs is within $\sim 2\%$ for the 
different cuts.

Furthermore, exclusion of the $Q^2 < 2$~GeV$^2$ region means a
reduction in the total number of data points by $\approx 50\%$,
since much of the existing polarized DIS data comes from experiments
performed at lower energies than unpolarized experiments.
Most global analyses of spin-dependent PDFs therefore choose
the $Q^2 \geq 1$~GeV$^2$ cut as a practical necessity.
As an alternative strategy, the NNPDF Collaboration \cite{NNPDF13}
use data down to $Q^2 = 1$~GeV$^2$, but impose a more stringent cut
of $W^2 > 6.25$~GeV$^2$ in the expectation that this will allow
for an analysis in terms of leading twist contributions only.
The impact of this cut on the JAM analysis appears from
Fig.~\ref{fig:cuts} to be more dramatic, especially for the
$\Delta d^+$ distribution, which at the $Q^2=1$~GeV$^2$ scale
is reduced by $\approx 15\%$ at $x \approx 0.5$, and by 20\%
at $x \approx 0.7$.
Such a large difference, together with the results found in
Fig.~\ref{fig:finQ2}(b), reaffirms the necessity of higher
twist corrections in analyses which utilize data down to
$Q^2 \approx 1$~GeV$^2$, as well as the need for new data at
higher $Q^2$ which can extend the constraints further into
the high-$x$ region.

\begin{figure}[t]
\includegraphics[width=8cm]{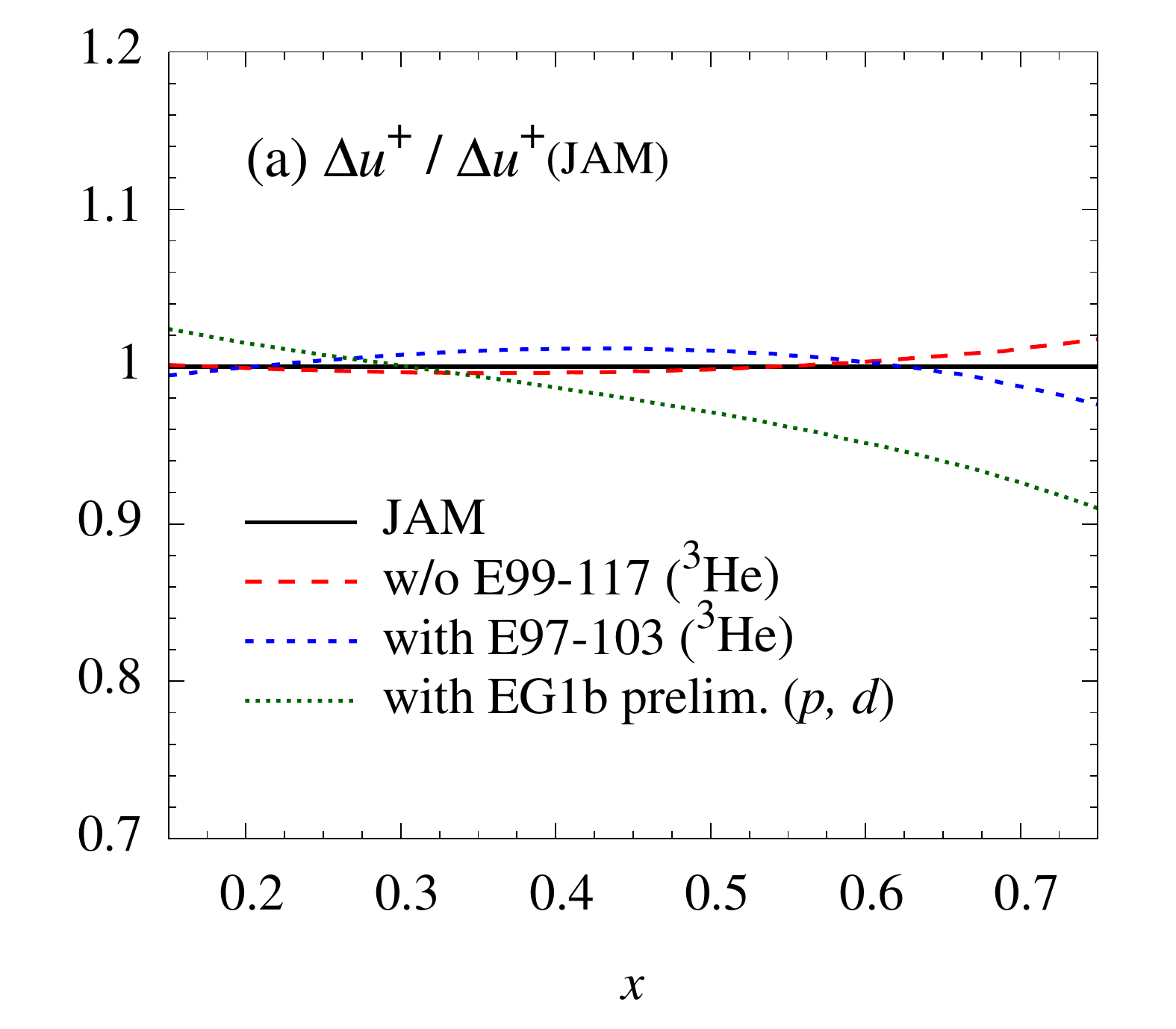}
\includegraphics[width=8cm]{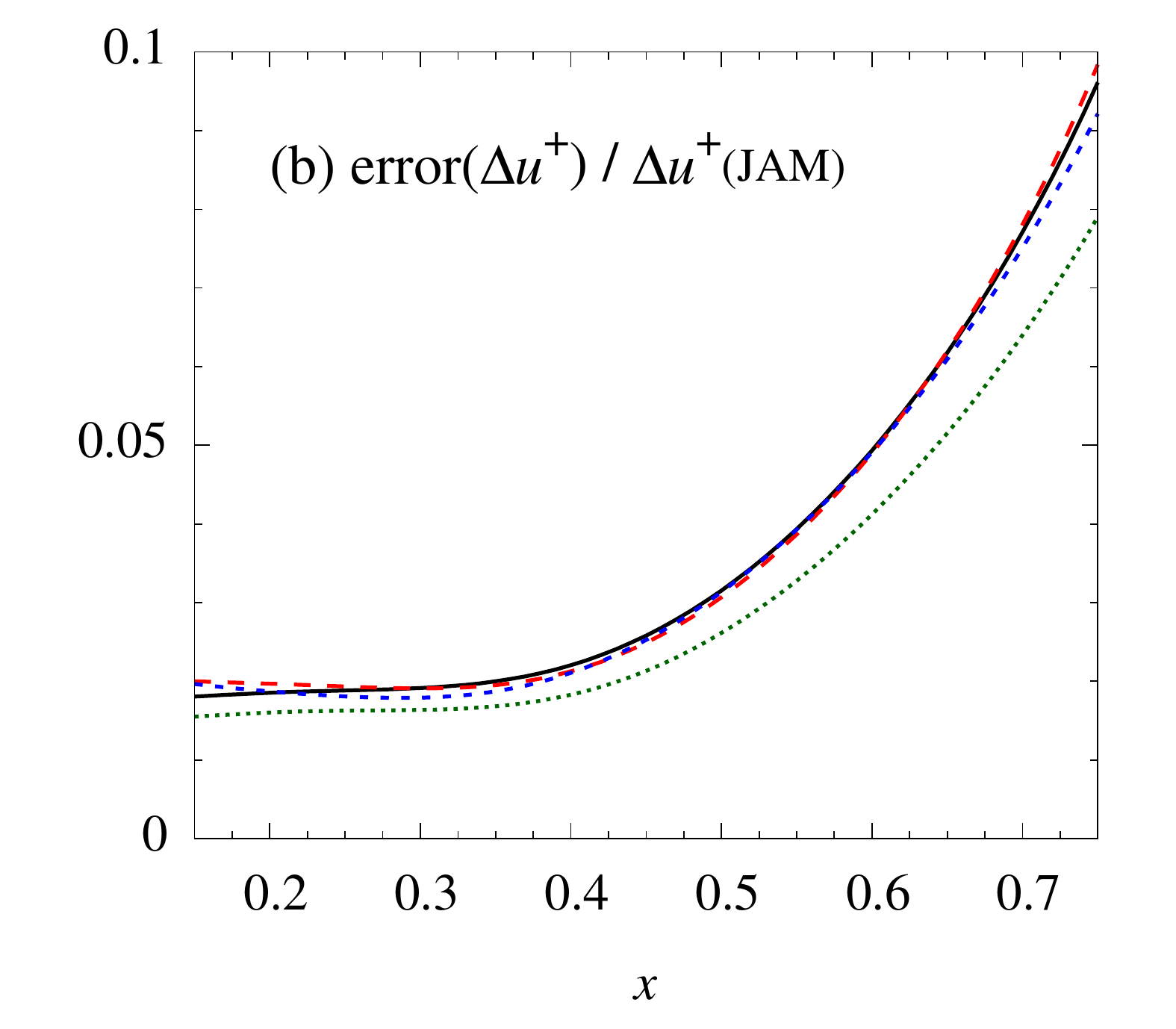}\\
\includegraphics[width=8cm]{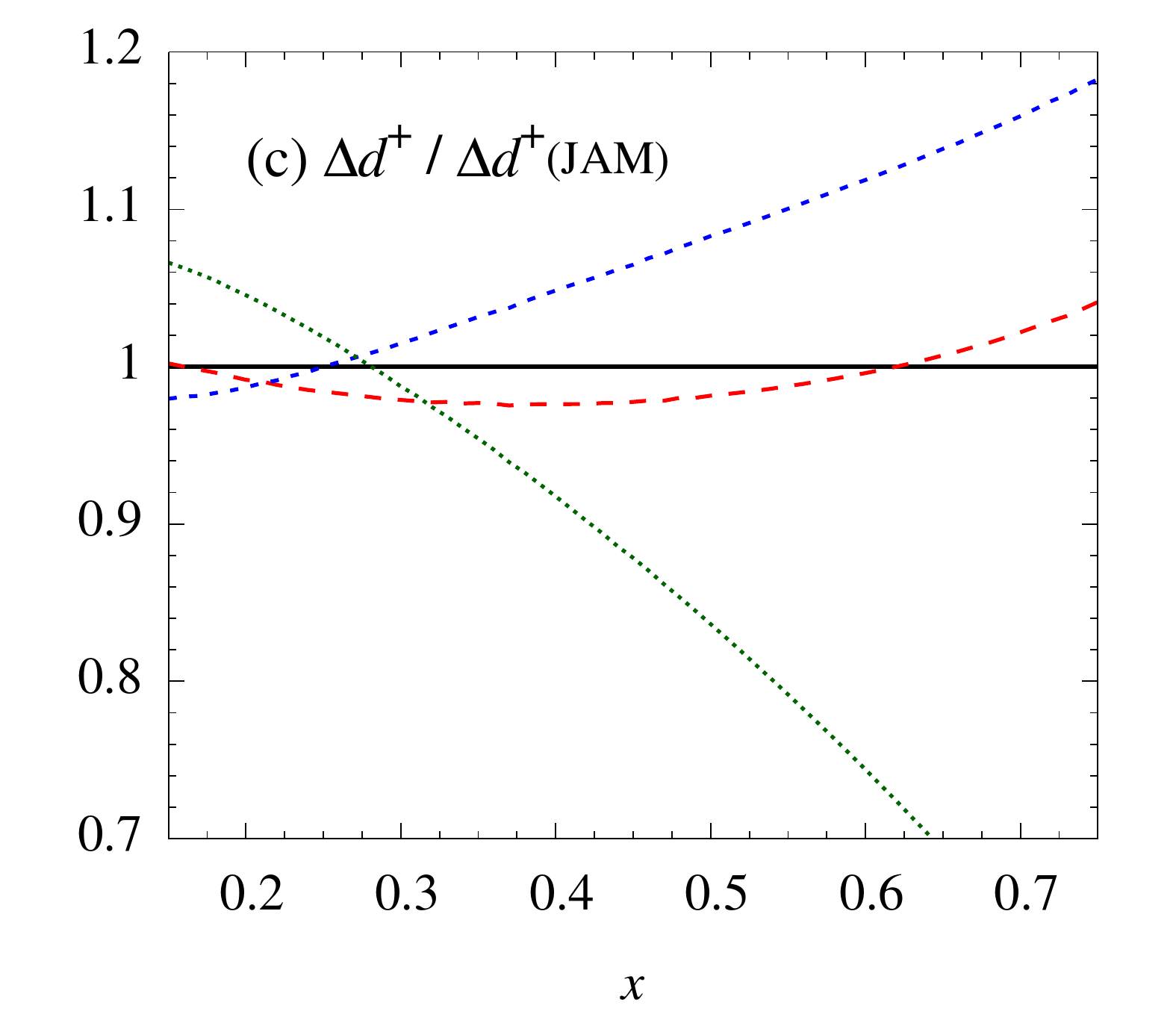}
\includegraphics[width=8cm]{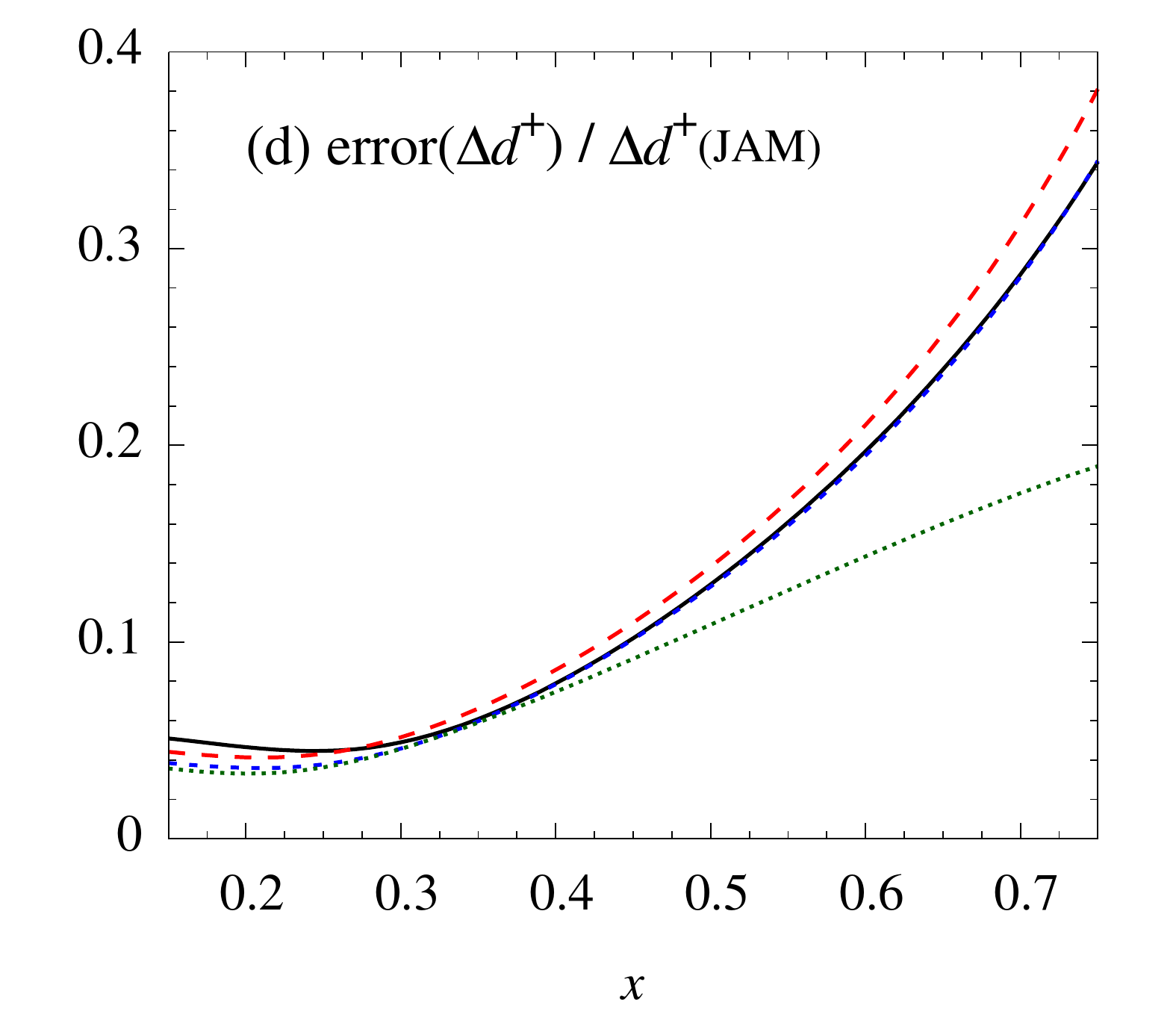}
\caption{Effects of various data sets on the $\Delta u^+$ and
	$\Delta d^+$ distributions and their errors, for the
	full JAM fit at $Q^2 = 1$~GeV$^2$ with the data sets
	in Table~\ref{tab:data} (black solid), compared with the
	fit without the Hall~A $^3$He data in Ref.~\cite{E99-117}
	(red dashed),
	with the Hall~A $^3$He data in Ref.~\cite{E97-103} included
	(blue short-dashed), and with the preliminary CLAS $p$ and $d$
	data in Ref.~\cite{EG1b} included.
	{\bf (a)} The ratio of the spectator $\Delta u^+$
	distribution to the JAM PDF, together with
	{\bf (b)} the relative error with respect to
	the central values of the JAM PDFs.
	The corresponding ratios for the $\Delta d^+$ PDFs
	are shown in {\bf (c)} and {\bf (d)}.}
\label{fig:JLab}
\end{figure}

To explore the impact of high-$x$, low-$W$ data on the global PDF
analysis, we examine the effect on the spin-dependent PDFs of adding
or removing specific data sets.  Since the Jefferson Lab Hall~A
experiment E99-117 \cite{E99-117} provided the most precise data
on the $^3$He polarization asymmetries at medium to large $x$
values (up to an average $\langle x \rangle = 0.6$), we compare
the results of the JAM fit, which includes these data, with
those obtained by excluding this experiment from the data set.
Surprisingly, Fig.~\ref{fig:JLab} indicates that there is almost
no difference in the central values of the fitted $\Delta u^+$
and $\Delta d^+$ distributions with or without the E99-117 data.
There is, however, a visible reduction of the error on the
$\Delta d^+$ distribution for $x \gtrsim 0.3$ with the inclusion
of the E99-117 data, by some $20\%-25\%$ at $x = 0.6-0.7$.

The E97-103 experiment \cite{E97-103} in Jefferson Lab Hall~A also
measured the $^3$He polarization asymmetries to very high precision,
at an average $x \approx 0.2$ and $Q^2 \lesssim 1.3$~GeV$^2$.
Although the number of additional data points from this experiment
is small (as with the E99-117 experiment), their errors are tiny,
which enables them to have a significant impact on the fits in
regions that have scant data or where the errors are large.
As expected, the effect of the E97-103 data is insignificant for
the $\Delta u^+$ distribution, but gives rise to a $\Delta d^+$
that is $\sim 10\%-20\%$ larger for $x \gtrsim 0.4$.
This is clearly an indirect effect of the fits, as the E97-103
data do not constrain this region directly.  However, it does
reduce the uncertainty on the $\Delta d^+$ PDF by some 30\%
at $x = 0.2$ compared with the main JAM fit.
Even more strikingly, the E97-103 data places strong constraints
on the higher twist part of $g_2^n$, pinching the error band at
$x \approx 0.2$, where the data are taken.  It is in fact not
possible to fit these data without inclusion of higher twist
contributions; doing so would lead to a 50\% reduction in the
$\Delta d^+$ distribution, in disagreement with other data.

In contrast, the new data (currently still being finalized) from the
EG1b experiment by the CLAS Collaboration at Jefferson Lab \cite{EG1b}
contribute over 760 points for $p$ and $d$ over a large kinematic
range, up to $Q^2 \approx 6.5$~GeV$^2$, albeit with larger errors
than for the Hall~A $^3$He data.
Inclusion of these points leads to a small suppression of the
$\Delta u^+$ distribution, which increases at higher $x$, but
a significant reduction in $\Delta d^+$, by up to $30\%-50\%$
for $x$ between $\approx 0.5$ and 0.7.
Concurrently, it also reduces the uncertainty on $\Delta u^+$
by $15\%-20\%$ in the large-$x$ region ($x \gtrsim 0.5$),
and even more significantly for the $\Delta d^+$, reducing it
by $\approx 30\%$ at $x=0.6$ and $\approx 50\%$ at $x=0.7$.
This clearly demonstrates the potential impact of the CLAS data
for reducing the overall PDF uncertainty at large $x$ values,
and will be vital to incorporate in future PDF analyses,
once the data analysis is finalized.

%%%%%%%%%%%%%%%%%%%%%%%%%%%%%%%%%%%%%%%%%%%%%%%%%%%%%%%%%%%%%%%%%%%%%%%%%
\section{Conclusion}
\label{sec:conc}

We have presented the first results from the JAM global NLO analysis
of spin-dependent PDFs from available data on inclusive polarized
DIS from protons, deuterons and $^3$He.
Where possible, we have fitted directly the measured polarization
asymmetries, rather than relying on structure functions extracted
under different conditions from the unpolarized cross sections.
We include data from all polarized DIS experiments that lie within
the limits $Q^2 \geq 1$~GeV$^2$ and $W^2 \geq 3.5$~GeV$^2$, which
allows us to constrain the $\Delta u^+$ and $\Delta d^+$ distributions
up to $x \approx 0.7$.  Obtaining stable fits over this expanded
kinematic range necessitates systematically accounting for target
mass and higher twist corrections, which are vital for describing
the $g_1$ and $g_2$ structure functions at the lower $Q^2$ range,
and nuclear smearing corrections for deuterium and $^3$He nuclei,
which have major impact at large~$x$.

The results of the main JAM fit indicate that the $\Delta d^+$
distribution has a significantly larger magnitude in the
intermediate-$x$ region ($x \gtrsim 0.2$) than in previous analyses,
due primarily to the sizeable higher twist corrections found here.
In particular, the twist $\tau=3$ term makes important contributions
to both the $g_1$ and $g_2$ structure functions of the proton, and
the $\tau=4$ correction makes a large and positive contribution to
the neutron $g_1$.  The latter is mostly responsible for driving
the $\Delta d^+$ distribution to become more negative.
The induced twist-3 contribution to the proton $g_1$ also reduces
the size of the twist-4 term compared to that found previously.
The $\tau=3$ correction to the neutron $g_2$ is compatible with
zero within errors.

The general features of the twist-3 corrections appear to be weakly
dependent on the choice of parametrization, except in the small-$x$
region, where some residual differences in their contributions to the
$g_2$ structure function are found.  In particular, parametrizations
that do not enforce the Burkhardt-Cottingham sum rule yield negative
(positive) enhancements at $x \lesssim 0.1$ for the proton (neutron),
leading to small violations of the sum rule.
On the other hand, differences in the behavior of the structure
functions at small $x$ are suppressed for higher moments, such as
the $d_2$ matrix element.  Our global analysis finds a positive
contribution to the proton $d_2$, of the order $\sim 0.01$ at a
scale $Q^2 = 5$~GeV$^2$, while the neutron $d_2$ is consistent
with zero.  Upcoming data from Jefferson Lab for $^3$He
\cite{E06-014} will better constrain the neutron $d_2$.
Independent of the details of the extracted higher twists,
our analysis clearly highlights the importance of including
subleading $1/Q^2$ corrections in any analysis that attempts
to fit data down to $Q^2 \sim 1$~GeV$^2$, even if more stringent
cuts in $W^2$ are imposed.  In contrast, relatively mild effects
are found when varying the $Q^2$ and $W^2$ cuts in the full JAM
analysis including finite-$Q^2$ and nuclear smearing corrections.

Future data \cite{12GeV} will also provide additional constraints
on the behavior of the ratios $\Delta q^+/q^+$ of polarized to
unpolarized PDFs in the $x \to 1$ limit.  This will be important
for testing perturbative QCD predictions for the $x \to 1$
behavior of PDFs, and for exploring the role of orbital angular
momentum.  While perturbative QCD arguments suggest that the
polarized to unpolarized ratio should asymptote to unity at
$x=1$ for all quark flavors, our fits for $\Delta d^+/d^+$
show no indication of a rise from its negative value over the
currently measured region.

Finally, while the present inclusive DIS data do not substantially
constrain the polarized sea quark and gluon distributions at small $x$,
in the next phase of the JAM analysis we shall integrate the lessons
learned here into an expanded study including semi-inclusive DIS
data and hadron production asymmetries in polarized $pp$ scattering,
as well as the remaining high-precision data from the completed
6~GeV experiments at Jefferson Lab \cite{EG1b}.
This will allow a more robust determination of the polarization of
the nucleon sea, and provide a baseline fit which can fully exploit
future data from Jefferson Lab at 12~GeV, RHIC, the EIC and elsewhere.

\newpage
%%%%%%%%%%%%%%%%%%%%%%%%%%%%%%%%%%%%%%%%%%%%%%%%%%%%%%%%%%%%%%%%%%%%%%%%%
\acknowledgments

We thank J.~Bl\"umlein, H.~B\"ottcher, V.~Braun, J.-P.~Chen,
C.~E.~Keppel, S.~Kuhn, S.~Kumano, E.~Nocera, O.~Rondon, B.~Sawatzky
and D.~Stamenov for helpful comments and discussions.
This work was supported by the DOE contract No.~DE-AC05-06OR23177,
under which Jefferson Science Associates, LLC operates Jefferson Lab.
The work of A.A. was supported in part by DOE contract No.~DE-SC0008791.

%%%%%%%%%%%%%%%%%%%%%%%%%%%%%%%%%%%%%%%%%%%%%%%%%%%%%%%%%%%%%%%%%%%%%%%%%


\begin{thebibliography}{99}

\bibitem{EMC89}
J.~Ashman {\it et al.},
Nucl. Phys. {\bf B328}, 1 (1989).

\bibitem{Kuhn09}
S.~E.~Kuhn, J.-P.~Chen and E.~Leader,
Prog. Part. Nucl. Phys. {\bf 63}, 1 (2009).

\bibitem{Aidala12}
C.~A.~Aidala, S.~D.~Bass, D.~Hasch and G.~K.~Mallot,
Rev. Mod. Phys. {\bf 85}, 655 (2013).

\bibitem{PJD13}
P.~Jimenez-Delgado, W.~Melnitchouk and J.~F.~Owens,
J. Phys. G: Nucl. Part. Phys. {\bf 40}, 093102 (2013).

\bibitem{LL13}
E.~Leader and C.~Lorc\'e,
arXiv:1309.4235.

\bibitem{DSSV09}
D.~de Florian, R.~Sassot, M.~Stratmann and W.~Vogelsang,
Phys. Rev. D {\bf 80}, 034030 (2009).

\bibitem{LSS10}
E.~Leader, A.~V.~Sidorov and D.~B.~Stamenov,
Phys. Rev. D {\bf 82}, 114018 (2010).

\bibitem{BB10}
J.~Bl\"umlein and H.~B\"ottcher,
Nucl. Phys. {\bf B841}, 205 (2010).

\bibitem{AAC09}
M.~Hirai and S.~Kumano,
Nucl. Phys. {\bf B813}, 106 (2009).

\bibitem{NNPDF13}
R.~D.~Ball {\it et al.},
Nucl. Phys. {\bf B874}, 36 (2013).

\bibitem{BSB02}
C.~Bourrely, J.~Soffer and F.~Buccella,
Eur. Phys. J. C {\bf 23}, 487 (2002).

\bibitem{AKS13} 
F.~Arbabifar, A.~N.~Khorramian and M.~Soleymaninia,
arXiv:1311.1830 [hep-ph].

\bibitem{MT96}
W.~Melnitchouk and A.~W.~Thomas,
Phys. Lett. B {\bf 377}, 11 (1996).

\bibitem{Isgur99}
N.~Isgur,
Phys. Rev. D {\bf 59}, 034013 (1999).

\bibitem{Holt10}
R.~J.~Holt and C.~D.~Roberts,  
Rev. Mod. Phys. {\bf 82}, 2991 (2010).

\bibitem{Avakian07}
H.~Avakian, S.~J.~Brodsky, A.~Deur and F.~Yuan,
Phys. Rev. Lett. {\bf 99}, 082001 (2007)

\bibitem{Chen11}
J.~P.~Chen, A.~Deur, S.~Kuhn and Z.-E.~Meziani,
J. Phys. Conf. Ser. {\bf 299}, 012005 (2011).

\bibitem{12GeV}
Jefferson Lab Experiments PR12-06-110,  % Hall C
J.-P.~Chen {\it et al.},
% X.~Zheng, Z.-E.~Meziani and G.~D.~Cates, spokespersons,
%
and PR12-06-122,			% Hall A
B.~Wojtsekhowski {\it et al.}, spokespersons.
% G.~Cates, N.~Liyanage, \mbox{Z.-E.~Meziani},
% G.~Rosner and X.~Zheng, spokespersons.

\bibitem{JAMweb}
The Jefferson Lab Angular Momentum (JAM) Collaboration website,\\
\url{http://www.jlab.org/JAM}.

\bibitem{CJ12} 
J.~F.~Owens, A.~Accardi and W.~Melnitchouk,
Phys. Rev. D {\bf 87}, 094012 (2013).

\bibitem{Ethier13} 
J.~J.~Ethier and W.~Melnitchouk,
Phys. Rev. C {\bf 88}, 054001 (2013).

\bibitem{CJ10}
A.~Accardi, M.~E.~Christy, C.~E.~Keppel, P.~Monaghan, W.~Melnitchouk,
J.~G.~Morfin and J.~F.~Owens,
Phys. Rev. D {\bf 81}, 034016 (2010).

\bibitem{CJ11}
A.~Accardi, W.~Melnitchouk, J.~F.~Owens, M.~E.~Christy, C.~E.~Keppel,
L.~Zhu and J.~G.~Morfin,
Phys. Rev. D {\bf 84}, 014008 (2011).

\bibitem{ABKM09}
S.~Alekhin, J.~Bl\"umlein, S.~Klein and S.-O.~Moch,
Phys. Rev. D {\bf 81}, 014032 (2010).

\bibitem{ABM12}
S.~Alekhin, J.~Bl\"umlein and S.-O.~Moch,
Phys. Rev. D {\bf 86}, 054009 (2012).

\bibitem{MMSTWW13} 
A.~D.~Martin, A.~J.~T.~M.~Mathijssen, W.~J.~Stirling, R.~S.~Thorne,
B.~J.~A.~Watt and G.~Watt,
Eur. Phys. J. C {\bf 73}, 2318 (2013).

\bibitem{JR-new}
P.~Jimenez Delgado and E.~Reya,
in preparation.

\bibitem{KM08}
S.~A.~Kulagin and W.~Melnitchouk,
Phys. Rev. C {\bf 77}, 015210 (2008).

\bibitem{KM09}
S.~A.~Kulagin and W.~Melnitchouk,
Phys. Rev. C {\bf 78}, 065203 (2008).

\bibitem{SMC98}
B.~Adeva {\it et al.},
Phys. Rev. D {\bf 58}, 112001 (1998).

\bibitem{SMC99}
B.~Adeva {\it et al.},
Phys. Rev. D {\bf 60}, 072004 (1999);
Erratum-ibid. D {\bf 62}, 079902 (2000).

\bibitem{COMPASS10}
M.~G.~Alekseev {\it et al.},
Phys. Lett. B {\bf 690}, 466 (2010).

\bibitem{COMPASS07}
V.~Yu.~Alexakhin {\it et al.},
Phys. Lett. B {\bf 647}, 8 (2007).

\bibitem{SLAC-E130}
G.~Baum {\it et al.},
Phys. Rev. Lett. {\bf 51}, 1135 (1983).

\bibitem{SLAC-E142}
P.~L.~Anthony {\it et al.},
Phys. Rev. D {\bf 54}, 6620 (1996).

\bibitem{SLAC-E143}
K.~Abe {\it et al.},
Phys. Rev. D {\bf 58}, 112003 (1998).

\bibitem{SLAC-E154}
K.~Abe {\it et al.},
Phys. Rev. Lett. {\bf 79}, 26 (1997);
%
Yu.~Kolomensky, Ph.D. thesis, U. Massachusetts (1997), SLAC-Report-503.

\bibitem{SLAC-E155p}
P.~L.~Anthony {\it et al.},
Phys. Lett. B {\bf 493}, 19 (2000).

\bibitem{SLAC-E155_A2pd}
P.~L.~Anthony {\it et al.},
Phys. Lett. B {\bf 458}, 529 (1999).

\bibitem{SLAC-E155d}                                            
P.~L.~Anthony {\it et al.},
Phys. Lett. B {\bf 463}, 339 (1999).

\bibitem{SLAC-E155x}
P.~L.~Anthony {\it et al.},
Phys. Lett. B {\bf 553}, 18 (2003).

\bibitem{HERMES07}
A.~Airapetian {\it et al.},
Phys. Rev. D {\bf 75}, 012007 (2007).

\bibitem{HERMES97}
K.~Ackerstaff {\it et al.},
Phys. Lett. B {\bf 404}, 383 (1997).

\bibitem{HERMES12}
A.~Airapetian {\it et al.},
Eur. Phys. J. C {\bf 72}, 1921 (2012).

\bibitem{E99-117}
X.~Zheng {\it et al.},
Phys. Rev. Lett. {\bf 92}, 012004 (2004);
Phys. Rev. C {\bf 70}, 065207 (2004).

\bibitem{E97-103}
K.~Kramer {\it et al.},
Phys. Rev. Lett. {\bf 95}, 142002 (2005);
K.~Kramer, Ph.D. thesis, College of William and Mary (2003).

\bibitem{EG1a}
K.~V.~Dharmwardane {\it et al.},
Phys. Lett. B {\bf 641}, 11 (2006).

\bibitem{EG1b}
Y.~Prok {\it et al.},
Phys. Lett. B {\bf 672}, 12 (2009);
%
S.~Kuhn,
private communication.

\bibitem{COMPASSg}
C.~Adolph {\it et al.},
Phys. Rev. D {\bf 87}, 052018 (2013).

\bibitem{JAMdb}
B.~Sawatzky {\it et al.},
The Jefferson Lab Angular Momentum (JAM) collaboration database,\\
\url{http://www.jlab.org/JAMDataBase}.

\bibitem{JAM-III}
P.~Jimenez-Delgado {\it et al.},
in preparation.

\bibitem{Lampe98} 
B.~Lampe and E.~Reya,
Phys. Rep. {\bf 332}, 1 (2000).

\bibitem{Weigl96}
T.~Weigl and W.~Melnitchouk,
Nucl. Phys. {\bf B465}, 267 (1996).

\bibitem{Wandzura77} 
S.~Wandzura and F.~Wilczek,
Phys. Lett. B {\bf 72}, 195 (1977).

\bibitem{BC70} 
H.~Burkhardt and W.~N.~Cottingham,
Annals Phys. {\bf 56}, 453 (1970).

\bibitem{JR09}
P.~Jimenez-Delgado and E.~Reya,
Phys. Rev. D {\bf 80}, 114011 (2009).

\bibitem{LSS98} 
E.~Leader, A.~V.~Sidorov and D.~B.~Stamenov,
Phys. Rev. D {\bf 58}, 114028 (1998).

\bibitem{Blankenbecler:1974tm}
R.~Blankenbecler and S.~J.~Brodsky,
Phys. Rev. D {\bf 10}, 2973 (1974).

\bibitem{Farrar:1975yb}
G.~R.~Farrar and D.~R.~Jackson,
Phys. Rev. Lett. {\bf 35}, 1416 (1975).

\bibitem{BBS95} 
S.~J.~Brodsky, M.~Burkardt and I.~Schmidt,
Nucl. Phys. {\bf B441}, 197 (1995).

\bibitem{JimenezDelgado:2012zx}
P.~Jimenez-Delgado,
Phys. Lett. B {\bf 714}, 301 (2012).

\bibitem{Pumplin:2001ct}
J.~Pumplin {\it et al.},
Phys. Rev. D {\bf 65}, 014013 (2001).

\bibitem{Aschenauer:2012ve} 
E.~C.~Aschenauer, R.~Sassot and M.~Stratmann,
Phys. Rev. D {\bf 86}, 054020 (2012).

\bibitem{Aschenauer:2013iia} 	% EIC
E.~C.~Aschenauer, T.~Burton, T.~Martini, H.~Spiesberger and M.~Stratmann,
arXiv:1309.5327 [hep-ph].

\bibitem{Accardi:2012qut}
A.~Accardi {\it et al.},
arXiv:1212.1701 [nucl-ex].
 
\bibitem{Accardi:2011mz}
A.~Accardi, V.~Guzey, A.~Prokudin and C.~Weiss,
Eur. Phys. J. A {\bf 48}, 92 (2012).

\bibitem{Adolph:2012ca}
C.~Adolph {\it et al.},		% [COMPASS Collaboration],
Phys. Rev. D {\bf 87}, 052018 (2013).

\bibitem{Adare:2008aa}          % A_LL(pi0)
A.~Adare {\it et al.},		% [PHENIX Collaboration],
Phys. Rev. Lett. {\bf 103}, 012003 (2009).

\bibitem{Adamczyk:2012qj}       % A_LL(jet)
L.~Adamczyk {\it et al.},	% [STAR Collaboration],
Phys. Rev. D {\bf 86}, 032006 (2012).

\bibitem{NNPDFeic}
R.~D.~Ball {\it et al.},
arXiv:1310.0461 [hep-ph].

\bibitem{MST94}
W.~Melnitchouk, A.~W.~Schreiber and A.~W.~Thomas,
Phys. Rev. D {\bf 49}, 1183 (1994);
Phys. Lett. B {\bf 335}, 11 (1994).

\bibitem{FS83}
L.~L.~Frankfurt and M.~I.~Strikman,
Nucl. Phys. {\bf A405}, 557 (1983).

\bibitem{Kaptari94}
L.~P.~Kaptari, K.~Yu.~Kazakov, A.~Yu.~Umnikov and B.~K\"ampfer,
Phys. Lett. B {\bf 321}, 271 (1994).

\bibitem{MPT95}
W.~Melnitchouk, G.~Piller and A.~W.~Thomas,
Phys. Lett. B {\bf 346}, 165 (1995); \\
%
G.~Piller, W.~Melnitchouk and A.~W.~Thomas,
Phys. Rev. C {\bf 54}, 894 (1996).

\bibitem{KMPW}
S.~A.~Kulagin, W.~Melnitchouk, G.~Piller and W.~Weise,
Phys. Rev. C {\bf 52}, 932 (1995).

\bibitem{Ciofi96}
C.~Ciofi degli Atti, L.~P.~Kaptari, S.~Scopetta and A.~Y.~Umnikov,
Phys. Lett. B {\bf 376}, 309 (1996).

\bibitem{Ciofi93} 
C.~Ciofi degli Atti, S.~Scopetta, E.~Pace and G.~Salm\`e,
Phys. Rev. C {\bf 48}, 968 (1993).

\bibitem{SS93}
R.-W.~Schulze and P.~U.~Sauer,
Phys. Rev. C {\bf 48}, 38 (1993);
Phys. Rev. C {\bf 56}, 2293 (1997).

\bibitem{Bissey01}
F.~R.~P.~Bissey, A.~W.~Thomas and I.~R.~Afnan,
Phys. Rev. C {\bf 64}, 024004 (2001).

\bibitem{BGST02}
F.~R.~P.~Bissey, V.~A.~Guzey, M.~Strikman and A.~W.~Thomas,
Phys. Rev. C {\bf 65}, 064317 (2002).

\bibitem{Forte13}
S.~Forte and G.~Watt,
Annu. Rev. Nucl. Part. Sci. {\bf 63}, 291 (2013).

\bibitem{Matsuda:1979ad}
S.~Matsuda and T.~Uematsu,
Nucl. Phys. {\bf B168}, 181 (1980).

\bibitem{Piccione:1997zh}
A.~Piccione and G.~Ridolfi,
Nucl. Phys. {\bf B513}, 301 (1998).

\bibitem{Blumlein:1998nv}
J.~Bl\"umlein and A.~Tkabladze,
Nucl. Phys. {\bf B553}, 427 (1999).

\bibitem{AM08}
A.~Accardi and W.~Melnitchouk,
Phys. Lett.  B {\bf 670}, 114 (2008).

\bibitem{BLMP11}
V.~M.~Braun, T.~Lautenschlager, A.~N.~Manashov and B.~Pirnay,
Phys. Rev. D {\bf 83}, 094023 (2011).

\bibitem{Braun01}
V.~M.~Braun, G.~P.~Korchemsky and A.~N.~Manashov,
Nucl. Phys. {\bf B603}, 69 (2001).

\bibitem{SGR}
G.~Sterman,
Nucl. Phys. {\bf B281}, 310 (1987);
%
S.~Catani and L.~Trentadue,
Nucl. Phys. {\bf B327}, 323 (1989);
%
S.~Schaefer, A.~Sch\"afer and M.~Stratmann,
Phys. Lett. B {\bf 514}, 284 (2001);
%
G.~Corcella and L.~Magnea,
Phys. Rev. D {\bf 72}, 074017 (2005).

\bibitem{JMC}
A.~Accardi and J.~W.~Qiu,
JHEP {\bf 07}, 090 (2008).

\bibitem{JMC13}
A.~Accardi and A.~Bacchetta,
in preparation.

\bibitem{ABMS09}
A.~Accardi, A.~Bacchetta, W.~Melnitchouk and M.~Schlegel,
JHEP {\bf 0911}, 093 (2009).

\bibitem{RSS07}
F.~R.~Wesselmann {\it et al.},
Phys. Rev. Lett. {\bf 98}, 132003 (2007).

\bibitem{RSS10} 
K.~Slifer {\it et al.},
Phys. Rev. Lett. {\bf 105}, 101601 (2010).

\bibitem{E01-012g2}
P.~Solvignon {\it et al.},
arXiv:1304.4497 [nucl-ex].

\bibitem{E06-014}
Jefferson Lab experiment E06-014,
S.~Choi {\it et al.}, spokespersons.
% X.~Jiang, \mbox{Z.-E.~Meziani} and B.~Sawatzky 

\bibitem{E01-012g1}
P.~Solvignon {\it et al.},
Phys. Rev. Lett. {\bf 101}, 182502 (2008).

\bibitem{BB12}
J.~Bl\"umlein and H.~B\"ottcher,
arXiv:1207.3170 [hep-ph].

\end{thebibliography}
\end{document}